\documentclass[aps,prb,twocolumn,showpacs,10pt,floatfix,superscriptaddress,floatfix]{revtex4-1}
\usepackage{braket}
\usepackage{dsfont}
\usepackage{amsfonts}
\usepackage{amsthm}
\usepackage{amssymb}
\usepackage{amsmath,amscd}
\usepackage{mathtools}
\usepackage{rotating}
\usepackage{color}
\usepackage{framed}
\usepackage{pbox}
\usepackage{xfrac}
\usepackage{extarrows}
\usepackage{enumerate}
\usepackage{graphicx}
\usepackage{float}
\usepackage{afterpage}
\usepackage{epigraph}
\usepackage{comment}
\usepackage[hidelinks]{hyperref}
\usepackage{mathrsfs}
\usepackage{tikz}
\usepackage{circuitikz}
\usepackage{tabularx}
\usepackage{ragged2e}
\usepackage[percent]{overpic}

\newcommand{\vk}{\vec k}
\newcommand{\ve}{\vec e}

\newcommand{\vj}{\vec j}

\newcommand{\vm}{\vec m}

\newcommand{\vmu}{\mbox{\boldmath $\mu$}}

\newcommand{\vh}{\vec h}

\newcommand{\Dex}{D_{\rm ex}}

\newcommand{\rrm}{\rho}

\newcommand{\vsigma}{\mbox{\boldmath $\sigma$}}
\newcommand{\vomega}{\mbox{\boldmath $\omega$}}
\newcommand{\vsigmascr}{\mbox{\scriptsize \boldmath $\sigma$}}
\newcommand{\vomegascr}{\mbox{\scriptsize \boldmath $\omega$}}

\renewcommand{\vec}[1]{\mathbf{#1}}

\begin{document}
\title{Finite-frequency spin conductance of a ferro-/ferrimagnetic-insulator$|$normal-metal interface}

\author{David A. Reiss}
\affiliation{Dahlem Center for Complex Quantum Systems and Physics Department, Freie Universit\"at Berlin, Arnimallee 14, 14195 Berlin, Germany}
\author{Piet W. Brouwer}
\affiliation{Dahlem Center for Complex Quantum Systems and Physics Department, Freie Universit\"at Berlin, Arnimallee 14, 14195 Berlin, Germany}

\begin{abstract}
The interface between a ferro-/ferrimagnetic insulator and a normal metal can support spin currents polarized collinear with and perpendicular to the magnetization direction. The flow of angular momentum perpendicular to the magnetization direction (``transverse'' spin current) takes place via spin torque and spin pumping. The flow of angular momentum collinear with the magnetization (``longitudinal'' spin current) requires the excitation of magnons. In this article we extend the existing theory of longitudinal spin transport [Bender and Tserkovnyak, Phys. Rev. B {\bf 91}, 140402(R) (2015)] in the zero-frequency weak-coupling limit in two directions: We calculate the longitudinal spin conductance non-perturbatively (but in the low-frequency limit) and at finite frequency (but in the limit of low interface transparency). For the paradigmatic spintronic material system YIG$|$Pt, we find that non-perturbative effects lead to a longitudinal spin conductance that is ca. 40\% smaller than the perturbative limit, whereas finite-frequency corrections are relevant at low temperatures $\lesssim 100\, {\rm K}$ only, when only few magnon modes are thermally occupied.
\end{abstract}

\maketitle

\section{Introduction}

In magnetic insulators, transport of angular momentum is possible via spin waves, collective wave-like excursions of the magnetization from its equilibrium direction.\cite{Ashcroft_Mermin,Kittel_2005,Landau_1980} A spin wave --- or its quantized counterpart, a ``magnon'' --- carries both an oscillating angular momentum current with polarization perpendicular (transverse) to and a non-oscillating angular momentum current with polarization parallel (longitudinal) to the magnetization direction. The magnitude of the transverse spin current is proportional to the amplitude of the spin wave; the magnitude of the longitudinal spin current is quadratic in the spin wave amplitude, {\em i.e.}, it scales proportional to the number of excited magnons.\cite{Gurevich_Melkov,Majlis_2007,Vonsovskii_1966} 

\begin{figure}
\centering
\includegraphics[width=0.45\textwidth]{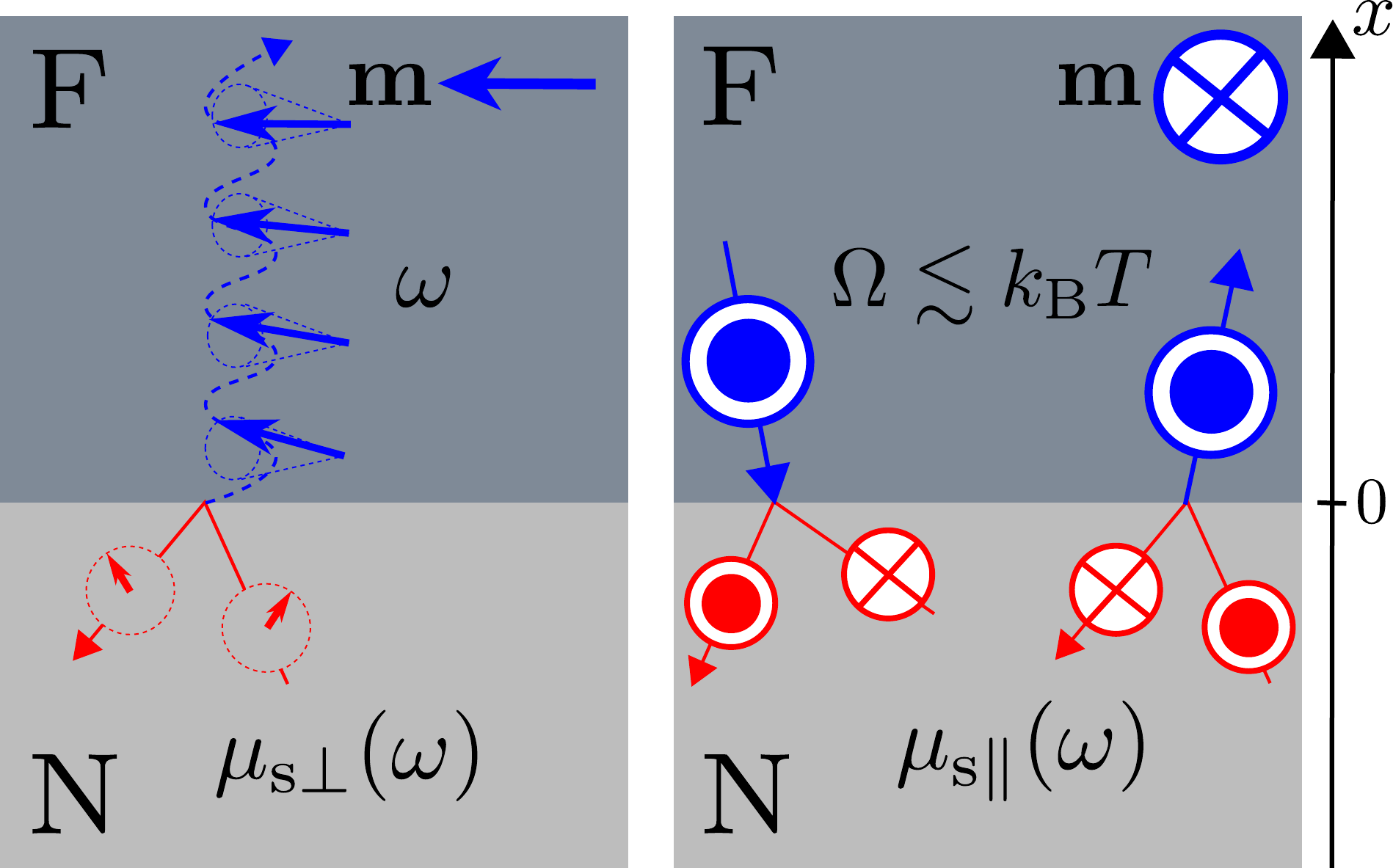}
\caption{Illustration of the microscopic mechanisms underlying the transverse (left) and longitudinal (right) components of the spin current through the interface between a ferro-/ferrimagnetic insulator F and a normal metal N. The transverse spin current is mediated by the spin torque and spin pumping involving electrons (red) with spins perpendicular to the magnetization and spin waves (blue) with frequency $\omega$ equal to the frequency at which the spin accumulation $\vmu_{\rm s}$ in N is driven. The longitudinal component arises from spin-flip scattering of conduction electrons (red), combined with the creation or absorption of thermal magnons of frequency $\Omega$ (blue). (The thermal magnon frequency $\Omega$ is not related to the driving frequency $\omega$.) Alternatively, the longitudinal component can be seen as arising from the spin torque exerted on/spin pumped by the transverse magnetization component induced by thermal fluctuations in F (not shown schematically).
 \label{fig:schematic} 
 }
\end{figure}

Both components of the spin current couple to conduction electrons at the interface between a ferro-/ferrimagnetic insulator (F) and a normal metal (N). Microscopically, the coupling of the transverse component can be understood in terms of the interfacial spin torque and spin pumping,\cite{Berger_1996,Slonczewski_1996,Slonczewski_1999,Tserkovnyak_2002_a,Tserkovnyak_2005} which both give an angular momentum current perpendicular to the magnetization direction, see Fig.\ \ref{fig:schematic} (left). A longitudinal spin current across the interface is obtained from the spin torque acting on or spin pumped by the small thermally-induced transverse magnetization component.\cite{Bender_2015} 
Alternatively and equivalently, a longitudinal interfacial spin current results from magnon-emitting or -absorbing scattering at the interface, as shown schematically in Fig.\ \ref{fig:schematic} (right). The transverse component of the interfacial spin current is relevant for coherent effects, such as the spin-torque diode effect\cite{Tulapurkar_2005,Sankey_2006} or the spin-torque induced ferromagnetic resonance.\cite{Liu_2011_2,Kondou_2012,Ganguly_2014} The longitudinal component governs incoherent effects, such as the interfacial contribution to the spin-Seebeck effect,\cite{Uchida-2008,Jaworski-2010,Uchida-2010,Bauer-2012} the spin-Peltier effect,\cite{Flipse_2014} or non-local magnonic spin-transport effects.\cite{Zhang_2012,Goennenwein_2015,Schlitz_2021} The spin-Hall magnetoresistance\cite{Weiler_2012,Huang_2012,Nakayama_2013,Hahn_2013,Vlietstra_2013,Althammer_2013} depends on a competition between both components of the spin current.\cite{Zhang_2019,Reiss_2021}

In the linear-response regime, the transverse spin current density $j^x_{{\rm s}\perp}$ through the F$|$N interface (directed from N to F) is proportional to the difference of the transverse spin accumulation $\mu_{{\rm s}\perp}$ in N and the time derivative of the transverse magnetization amplitude $m_{\perp}$ at the interface,\cite{Tserkovnyak_2005}
\begin{equation}
\label{eq:gperp}
  j^x_{{\rm s}\perp}(\omega) = \frac{g_{\uparrow\downarrow}}{4 \pi} [\mu_{{\rm s}\perp}(\omega) + \hbar \omega m_{\perp}(\omega)]. 
\end{equation}

\noindent
The coefficient of proportionality $g_{\uparrow\downarrow}$ is complex and known as the ``spin-mixing conductance'' per unit area.\cite{Brataas_2000}
Omitting Seebeck-type contributions that depend on the temperature difference across the F$|$N interface, the longitudinal spin current density $j^x_{{\rm s}\parallel}$ is proportional to the difference of the longitudinal spin accumulation $\mu_{{\rm s}\parallel}$ in N and the ``magnon chemical potential'' $\mu_{{\rm m}}$,\cite{Cornelissen_2016}
\begin{equation}
  j^x_{{\rm s}\parallel}(\omega) = \frac{g_{{\rm s}\parallel}}{4 \pi} [\mu_{{\rm s}\parallel}(\omega) - \mu_{{\rm m}}(\omega)].
  \label{eq:jparallel}
\end{equation}

\noindent
In the limit of weak coupling across the F$|$N interface the longitudinal interfacial spin conductance is proportional to the real part of the spin-mixing conductance,\cite{Bender_2015,Cornelissen_2016,Schmidt_2018}
\begin{equation}
  g_{{\rm s}\parallel} = \frac{4\, \mbox{Re}\, g_{\uparrow\downarrow}}{s}
  \int d\Omega \, \nu_{\rm m}(\Omega) \Omega \left( - \frac{df_T(\Omega)}{d\Omega} \right) \!.
  \label{eq:glong}
\end{equation}

\noindent
Here $s$ is the spin per volume in F, $\nu_{\rm m}(\Omega)$ the density of states (DOS) of magnon modes at frequency $\Omega$, and $f_T(\Omega)$ the Planck distribution at temperature $T$ of the magnons.

The availability of high-quality THz sources, combined with spin-orbit-mediated conversion of electric into magnetic driving, as well as of femtosecond laser pulses for pump-probe spectroscopy has made it possible to experimentally access spin transport across F$|$N interfaces on ultrafast time scales.\cite{Schellekens_2014,Razdolski_2017,Seifert_2018,Brandt_2021,Jimenez-Cavero_2022,Kimling_2017,Kholid_2021} 
Whereas Eq.\ (\ref{eq:gperp}) is valid for frequencies small in comparison to the frequencies of acoustic magnons at the zone boundary,\cite{Tserkovnyak_2002_a} which reach well into the THz regime, Eq.\ (\ref{eq:glong}) requires driving frequencies much smaller than the frequencies of {\em thermal} magnons, \textit{i.e.}, $\omega/2 \pi \lesssim k_{B} T/ h \approx 6.3\,$THz for $300\,$K.\cite{Bender_2015} At room temperature, the two conditions roughly coincide for the magnetic insulator YIG, which is the material of choice for many experiments, or for ferrites, such as CoFe$_2$O$_4$ and NiFe$_2$O$_4$, see Refs. \onlinecite{Teh_1973} and \onlinecite{Shan_2018}. But at low temperatures, the condition for the applicability of Eq.\ (\ref{eq:glong}) is stricter and may be violated for sufficiently fast driving for these materials.\footnote{The effects discussed here in principle arise in magnetic \textit{metals}, too. However, in that case these effects are difficult to be distinguished experimentally from the much larger spin conductance carried by electrons in the magnet.\cite{Reiss_2021} This is the reason why we restrict the present discussion to magnetic insulators.} An example of a magnetic material for which the two conditions do not coincide already at \textit{room temperature} is Fe$_3$O$_4$ (magnetite), for which the frequency of acoustic magnons at the zone boundary is well above the frequency of thermal magnons at room temperature.\cite{Krupicka_1982}

In this article, we present two calculations of the longitudinal interfacial spin conductance $g_{{\rm s}\parallel}(\omega)$ per area that go beyond the low-frequency weak-coupling regime of validity of Eq.\ (\ref{eq:glong}): (i) We calculate $g_{{\rm s}\parallel}$ in the low-frequency limit, but without the assumption of weak coupling across the F$|$N interface, and (ii) we calculate the finite-frequency longitudinal spin conductance $g_{{\rm s}\parallel}(\omega)$ per area in the weak-coupling limit. Our finite-frequency result is applicable in the same frequency range as Eq.\ (\ref{eq:gperp}), {\em i.e.}, within the entire frequency range of acoustic magnons. Additionally, the temperature $T$ must be low enough such that only acoustic magnons are thermally excited. For YIG this condition amounts to the requirement that $T \lesssim 300\,$K.\cite{Barker_2016} Comparing our non-perturbative low-frequency calculation to the weak-coupling result in Eq. (\ref{eq:glong}), we find that the latter is a good order-of-magnitude estimate for most material combinations, whereas quantitative deviations are possible. 

This article is organized as follows: In Sec.\ \ref{sec:2} we report our non-perturbative calculation of the longitudinal spin conductance $g_{{\rm s}\parallel}$ at zero frequency, using scattering theory for the reflection of spin waves from the F$|$N interface. In Sec. \ref{sec:3} we present our perturbative calculation of the finite-frequency longitudinal spin conductance $g_{{\rm s}\parallel}(\omega)$, using the method of non-equilibrium Green functions. We give numerical estimates for material combinations involving the magnetic insulator YIG in Sec.\ \ref{sec:4} and we conclude in Sec.\ \ref{sec:5}. Appendices \ref{app:a} and \ref{app:Keldysh} contain further details of the calculations.

\section{Non-perturbative calculation at zero frequency}
\label{sec:2}

Central to our non-perturbative calculation is the amplitude $\rrm (\Omega)$ that a magnon with frequency $\Omega$ incident on the F$|$N interface is reflected back into F. The ``transmission coefficient'' $|\tau(\Omega)|^2 = 1 - |\rrm(\Omega)|^2$ is the probability that the magnon is not reflected and, instead, transfers its angular momentum $\hbar$ to the conduction electrons in N. As we show below, knowledge of $\rrm(\Omega)$ is sufficient for the calculation of the longitudinal interfacial spin conductance $g_{{\rm s}\parallel}(\omega)$ per area in the low-frequency limit. 

{\em Magnon reflection amplitude $\rrm$.---}
To keep the notation simple, we describe our calculation for a one-dimensional geometry and switch to  three dimensions in the presentation of the final results. We consider an F$|$N interface with coordinate $x$ normal to the interface and a magnetic insulator F for $x > 0$, see Fig.\ \ref{fig:schematic}. Magnetization dynamics in F is described by the Landau-Lifshitz equation
\begin{equation}
  \dot \vm = \omega_0 \, \ve_{\parallel} \times \vm + \frac{1}{\hbar s} \frac{\partial}{\partial x} \vj_{\rm s}^x, \label{eq:llg}
\end{equation}

\noindent
where $\vm$ is a unit vector pointing along the direction of the magnetization, $\omega_0$ is the ferromagnetic resonance frequency, $\ve_{\parallel}$ the equilibrium magnetization direction, and
\begin{equation}
  \vj_{\rm s}^x = -\hbar s \Dex \vm \times \frac{\partial \vm}{\partial x}
  \label{eq:js}
\end{equation}

\noindent
the spin current density, with $\Dex$ the spin stiffness of dimension length$^2 \cdot$ time$^{-1}$. (We recall that the gyromagnetic ratio is negative, so that the angular momentum density corresponding to the magnetization direction $\vm$ is $-\hbar s \vm$.)
The spin current density through the F$|$N interface is\cite{Brataas_2000,Tserkovnyak_2005,Tserkovnyak_2002_a,Foros_2005,Tatara2017}
\begin{align}
  \label{eq:jvectorboundary}
  \vj^x_{\rm s} =&\, - \frac{1}{4 \pi}
    (\mbox{Re}\, g_{\uparrow\downarrow}\, \vm \times \mbox{} + \mbox{Im}\, g_{\uparrow\downarrow})
  \left[ (\vm \times \vmu_{\rm s} )  + \hbar \dot \vm \right]
  \nonumber \\ &\, \mbox{}
  + \hbar \sqrt{\frac{\mbox{Re}\, g_{\uparrow\downarrow}}{2 \pi}} \, \vm \times \vh' ,
\end{align}

\noindent
where $g_{\uparrow\downarrow}$ is the complex spin-mixing conductance\footnote{Equation (\ref{eq:jvectorboundary}) contains two terms proportional to the imaginary part of the spin mixing conductance, $- (1/4 \pi) \mbox{Im}\, g_{\uparrow\downarrow} (\vm \times \vmu_{\rm s})$ and $- (1/4 \pi) \mbox{Im}\, g_{\uparrow\downarrow} (\hbar \dot{\vm})$. In a microscopic theory of the F$|$N interface, the coefficient multiplying $\hbar \dot{\vm}$ may be different from the coefficient multiplying $\vm \times \vmu_{\rm s}$, see Ref.\ \onlinecite{Tatara2017}. Since the numerical values for the imaginary part of the spin mixing conductance used to estimate the longitudinal spin conductance in Sec.\ \ref{sec:4} are much smaller than those for its real part and the longitudinal spin conductance is mainly determined by $\mbox{Re}\, g_{\uparrow\downarrow}$, we ignore this difference here.} and $\vh'$ is proportional to a stochastic magnetic field representing the spin torque due to current fluctuations in N. If the normal metal is in equilibrium at temperature $T_{\rm N}$, the correlation function of the stochastic term $\vh'$ is given by the fluctuation-dissipation theorem,\cite{Foros_2005}
\begin{equation}
  \langle h'_{\alpha}(\Omega')^* h'_{\beta}(\Omega) \rangle = \Omega f_{T_{\rm N}}(\Omega) \delta(\Omega-\Omega') \delta_{\alpha\beta}, \label{eq:bTN}
\end{equation}

\noindent
where $f_T(\Omega) = 1/(e^{\hbar \Omega/k_{\rm B} T}-1)$ is the Planck function and the Fourier transform is defined as
\begin{equation}
  \vh'(t) = \frac{1}{\sqrt{2 \pi}} \int\limits_{-\infty}^{\infty} d\Omega\, \vh'(\Omega) e^{-i \Omega t}.
\end{equation}

We parameterize the magnetization direction $\vm$ as
\begin{align}
  \vm(x,t) =&\, \sqrt{1 - 2 |m_{\perp}(x,t)|^2} \, \ve_{\parallel} 
  \nonumber \\ &\, \mbox{}
  + m_{\perp}(x,t) \ve_{\perp} + m_{\perp}(x,t)^* \ve_{\perp}^*,
  \label{eq:mdecomp}
\end{align}

\noindent
where the complex unit vectors $\ve_{\perp}$ and $\ve_{\perp}^*$ span the directions orthogonal to the equilibrium magnetization direction $\ve_{\parallel}$ and satisfy the condition $\ve_{\perp} \times \ve_{\parallel} = i \ve_{\perp}$.
The solution of the Landau-Lifshitz equation (\ref{eq:llg}), up to linear order in the magnetization amplitude $m_{\perp}$, then reads 
\begin{align}
  m_{\perp}(x,t) =&\, 
  \int\limits_{-\infty}^{\infty} d\Omega \,
  \frac{e^{- i \Omega t}}{\sqrt{4 \pi s \Dex k_x}}
  \nonumber \\ &\, \mbox{} \times
  \left[a_{\rm in}(\Omega) e^{-i k_x x} + a_{\rm out}(\Omega) e^{i k_x x} \right] \!,
  \label{eq:mgeneral}
\end{align}

\noindent
where
\begin{equation}
  k_x(\Omega) = \sqrt{\frac{\Omega-\omega_0}{\Dex}} \label{eq:kx}
\end{equation}

\noindent
and $a_{\rm in}(\Omega)$ and $a_{\rm out}(\Omega)$ are flux-normalized amplitudes for spin waves moving towards the F$|$N interface at \mbox{$x=0$} and away from it, respectively. (The amplitudes $a_{\rm in}(\Omega)$ and $a_{\rm out}(\Omega)$ may be interpreted as magnon annihilation operators in a quantized formulation.)
The spin current density $\vj_{{\rm s}}^x$ can be decomposed into transverse and longitudinal contributions analogous to Eq.\ (\ref{eq:mdecomp}),
\begin{equation}
  \vj_{\rm s}^x(x,t) = j_{{\rm s}\parallel}^x(x,t) \ve_{\parallel} + j_{{\rm s}\perp}^x(x,t) \ve_{\perp} + j_{{\rm s}\perp}^{x}(x,t)^* \ve_{\perp}^*.
  \label{eq:jFN}
\end{equation}

\noindent
In the same way, the spin accumulation $\vmu_{\rm s}$ and the stochastic term $\vh'$ can be decomposed into transverse and longitudinal contributions.

We first consider the transverse spin current density $j_{{\rm s}\perp}^x$ to linear order in the magnetization amplitude $m_{\perp}$. From Eqs.\ (\ref{eq:js}) and (\ref{eq:mgeneral}), one finds that the magnonic transverse spin current density $j^x_{{\rm s}\perp}(0,t)$ at the F$|$N interface $x=0$ is
\begin{align}
       \label{eq:jsperp}
  j^x_{{\rm s}\perp}(0,t) =&\,
  i \hbar s \Dex \frac{\partial m_{\perp}(x,t)}{\partial x} \\ \nonumber  =&\,
  \frac{\hbar}{4 \pi}
  \int\limits_{-\infty}^{\infty} d\Omega \, e^{- i \Omega t} \sqrt{4 \pi s \Dex k_x(\Omega)}
  \nonumber \\ &\, \mbox{} \times
       [a_{\rm in}(\Omega) - a_{\rm out}(\Omega)]. \nonumber
\end{align}

\noindent
Equation (\ref{eq:jvectorboundary}) implies that the transverse spin current density through the interface is given by
\begin{align}
  j^x_{{\rm s}\perp}(0,t) =&\, \frac{g_{\uparrow\downarrow}}{4 \pi} 
  [\mu_{{\rm s}\perp}(t) + i \hbar \dot m_{\perp}(0,t)
    - \mu_{{\rm s}\parallel}(t) m_{\perp}(0,t)] \nonumber \\ &\, \mbox{}
  -i \hbar \sqrt{\frac{\mbox{Re}\, g_{\uparrow\downarrow}}{2 \pi}} h'_{\perp}(t) .
  \label{eq:jsperpboundary}
\end{align}

\noindent
Imposing continuity of the transverse spin current at the F$|$N interface allows us to express the amplitude $a_{\rm out}$ of magnons moving away from the interface in terms of the amplitude $a_{\rm in}$ of incident magnons and the stochastic field $h'_{\perp}$. 
Inserting Eqs.\ (\ref{eq:mgeneral}) and (\ref{eq:jsperp}) into the boundary condition (\ref{eq:jsperpboundary}), we get
%
\begin{align}
  a_{\rm out}(\Omega) =&\, \rrm(\Omega) a_{\rm in}(\Omega)
  + \rrm'(\Omega) h'_{\perp}(\Omega),
  \label{eq:aoutzero}
\end{align}

\noindent
with
\begin{align}
  \label{eq:rho}
  \rrm(\Omega) =&\, \frac{4 \pi s \Dex k_x(\Omega) - (\Omega - \mu_{{\rm s}\parallel}/\hbar) g_{\uparrow\downarrow}}{4 \pi s \Dex k_x(\Omega) + (\Omega - \mu_{{\rm s}\parallel}/\hbar) g_{\uparrow\downarrow}}, \nonumber \\
  \rrm'(\Omega) =&\, \frac{2 \sqrt{4 \pi s \Dex k_x(\Omega)\, \mbox{Re}\, g_{\uparrow\downarrow}}}{4 \pi s \Dex k_x(\Omega) + (\Omega - \mu_{{\rm s}\parallel}/\hbar) g_{\uparrow\downarrow}}.
\end{align}

\noindent
The coefficient $\rrm(\Omega)$ is the amplitude that a magnon with frequency $\Omega$ incident on the F$|$N interface is reflected. One therefore may interpret
\begin{align}
  |\tau(\Omega)|^2 =&\,
  1 - |\rrm(\Omega)|^2
  \nonumber \\ =&\,
  (\Omega - \mu_{{\rm s}\parallel}/\hbar)
  |\rrm'(\Omega)|^2
  \label{eq:tau}
\end{align}
as the probability that a magnon is annihilated at the F$|$N interface while exciting a spinful excitation in N.



{\em Longitudinal interfacial spin conductance.---} The longitudinal spin current is quadratic in the magnetization amplitude. From Eqs.\ (\ref{eq:js}) and (\ref{eq:jFN}) one finds
\begin{equation}
  j^x_{{\rm s}\parallel}(0,t) = m_{\perp}(0,t)^* j^x_{{\rm s}\perp}(0,t) +
  j^x_{{\rm s}\perp}(0,t)^* m_{\perp}(0,t),
\end{equation}

\noindent
so that continuity of $j^x_{{\rm s}\perp}$ at the F$|$N interface to linear order in $m_{\perp}$ also ensures continuity of $j^x_{{\rm s}\parallel}$. In terms of the magnon amplitudes, we find from Eqs.\ (\ref{eq:js}) and (\ref{eq:mgeneral}) that
\begin{align}
    \label{eq:jsxparallel_out_in}
  j^x_{{\rm s}\parallel}(0,t) =&\, \hbar \int\limits_{-\infty}^{\infty} \frac{d\omega}{2 \pi} e^{-i \omega t} \int\limits_{-\infty}^{\infty} d\Omega  \\ \nonumber &\, 
  \mbox{} \times
  [a_{\rm out}(\Omega_-)^* a_{\rm out}(\Omega_+)
   - a_{\rm in}(\Omega_-)^* a_{\rm in}(\Omega_+)],
\end{align}

\noindent
where we abbreviated $\Omega_{\pm} = \Omega \pm \omega/2$
and omitted terms that drop out in the limit $\omega \to 0$.
The correlation function of the magnon amplitudes is given by the (quantum-mechanical) fluctuation-dissipation theorem,\cite{Landau_1_1980}
\begin{equation}
  \langle a_{\rm in}(\Omega_-)^* a_{\rm in}(\Omega_+) \rangle 
  =\, f_{T_{\rm F}}(\Omega - \mu_{\rm m}/\hbar) \delta(\omega).
  \label{eq:acorrelatorzero}
\end{equation}

\noindent
Here $T_{\rm F}$ is the (magnon) temperature of the magnetic insulator and $f_{T_{\rm F}}(\Omega) = 1/(e^{\hbar \Omega/k_{\rm B} T_{\rm F}}-1) \Theta(\Omega-\omega_0)$ the Planck function, with $\Theta$ the Heaviside step function.
To obtain the correlation function of the stochastic field $\vh'$ in the presence of a spin accumulation $\vmu_{\rm s} = \mu_{\rm s \parallel} \ve_{\parallel}$, we use the equilibrium result in Eq. (\ref{eq:bTN}) and make use of the fact that a spin accumulation $\vmu_{\rm s}$ can be shifted away by transforming to a spin reference frame that rotates at angular frequency $\vomega = \vmu_{\rm s}/\hbar$, see App.\ \ref{app:a}.
Denoting the stochastic field in the rotating frame by $\tilde \vh'$, we then have
\begin{equation}
  \tilde h'_{\perp}(\Omega) = h'_{\perp}(\Omega + \mu_{{\rm s}\parallel}).
\end{equation}

\noindent
In the rotating frame there is no spin accumulation in N, so that the correlation function of $\tilde h'_{\perp}$ is given by Eq.\ (\ref{eq:bTN}). It follows that
\begin{align}
    \langle h'_{\perp}(\Omega_-)^* h'_{\perp}(\Omega_+) \rangle =&\,
   (\Omega - \mu_{{\rm s}\parallel}/\hbar) \nonumber \\ &\, \mbox{} \times f_{T_{\rm N}}(\Omega - \mu_{{\rm s}\parallel}/\hbar) \, \delta(\omega). \label{eq:bTN2}
\end{align}

\noindent
Inserting this result as well as Eqs. \eqref{eq:aoutzero}, \eqref{eq:tau}, and \eqref{eq:acorrelatorzero} into Eq. \eqref{eq:jsxparallel_out_in}, we find for the longitudinal spin current 
\begin{align}
  \label{eq:jszero}
  j^x_{{\rm s}\parallel} =&\, \frac{\hbar}{2 \pi}
  \int\limits_{\omega_0}^{\infty} d\Omega\, 
  |\tau(\Omega)|^2
  \nonumber \\ &\, \mbox{} \times
            [f_{T_{\rm N}}(\Omega - \mu_{{\rm s \parallel}}/\hbar) - f_{T_{\rm F}}(\Omega - \mu_{\rm m}/\hbar)].
\end{align}
%
\noindent
Equation (\ref{eq:jszero}), together with Eq.\ (\ref{eq:tau}) for $|\tau(\Omega)|^2$, illustrates the equivalence of the two pictures of longitudinal spin transport mentioned in the introduction: as arising from magnon-emitting/absorbing scattering at the F$|$N interface (see first line in Eq.\ (\ref{eq:tau})) as well as from stochastic spin torques due to thermal fluctuations (see second line in Eq.\ (\ref{eq:tau})).

{\em In three dimensions} the calculation of the longitudinal spin current density involves an integration over modes with transverse wavenumbers $(k_y,k_z)$. For each transverse mode the previous calculation applies, but with $k_x(\Omega)$ replaced by
\begin{equation}
  \label{eq:kx3d}
  k_x(\Omega, k_{\perp}) = \sqrt{\frac{\Omega - \omega_0}{\Dex} - k_{\perp}^2},
\end{equation}
with $k_{\perp}^2 = k_y^2+k_z^2$. In particular, the mode-dependent reflection amplitude $\rho(\Omega,k_{\perp})$ and transmission coefficient $|\tau(\Omega,k_{\perp})|^2$ are found by substituting $k_x(\Omega, k_{\perp})$ for $k_x(\Omega)$ in Eq.\ (\ref{eq:rho}). For the steady-state longitudinal spin current density we then find
\begin{align}
j^x_{{\rm s}\parallel}
=&\, \frac{\hbar}{2(2 \pi)^2}
\int\limits_{\omega_0}^{\infty} d\Omega\, 
  k_x(\Omega)^2 T_{\rm m} (\Omega)
\nonumber \\ 
&\, \mbox{} \times [f_{T_{\rm N}}(\Omega - \mu_{{\rm s} \parallel}/\hbar) - f_{T_{\rm F}}(\Omega - \mu_{\rm m}/\hbar)],
\label{eq:jszero3d}
\end{align}

\noindent
where $k_x(\Omega)$ is given by Eq.\ (\ref{eq:kx}) and
$T_{\rm m} (\Omega)$ is the mode-averaged magnon transmission coefficient,
\begin{align}
  T_{\rm m} (\Omega) = \frac{2}{k_x(\Omega)^2} \int\limits_{0}^{k_x(\Omega)} dk_{\perp} \, k_{\perp} |\tau(\Omega,k_{\perp})|^2.
  \label{eq:Tm}
\end{align}

The validity of Eqs.\ (\ref{eq:jszero}) and (\ref{eq:jszero3d}) is not restricted to linear response or to weak coupling across the F$|$N interface. For comparison with the literature and with the perturbative calculation of the next section, it is nevertheless instructive to expand Eqs.\ (\ref{eq:jszero}) and (\ref{eq:jszero3d}) to linear order in the interfacial spin-mixing conductance, which gives
\begin{align}
  \label{eq:jsperturbative}
  j^x_{{\rm s}\parallel} =&\,
  \frac{1}{\pi s} \mbox{Re}\, g_{\uparrow\downarrow}
  \int\limits_{\omega_0}^{\infty} d\Omega\, \nu_{\rm m}(\Omega)
  (\hbar \Omega - \mu_{{\rm s}\parallel}) 
  \nonumber \\ &\, \mbox{} \times
  [f_{T_{\rm N}}(\Omega - \mu_{{\rm s}\parallel}/\hbar) - f_{T_{\rm F}}(\Omega - \mu_{\rm m}/\hbar)],
\end{align}

\noindent
where $\nu_{\rm m}(\Omega)$ is the magnon density of states, which equals $\nu_{\rm m}^{\rm 1D}(\Omega) = 1/2 \pi \Dex k_x(\Omega)$ in the one-dimensional case and $\nu^{\rm 3D}_{\rm m}(\Omega) = k_x(\Omega)/4 \pi^2 \Dex$ in the three-dimensional case. One verifies that this expression is consistent with Eq.\ (\ref{eq:glong}) to linear order in $\mu_{\rm s \parallel}- \mu_{\rm m}$.

\section{Perturbative calculation at finite frequencies} 
\label{sec:3}

In this section we again consider the longitudinal spin current density $j_{{\rm s}\parallel}^x$ through the interface between a ferro-/ferrimagnetic insulator F and a normal metal N, but now with a time-dependent spin accumulation $\mu_{{\rm s}\parallel}(t)$ in N. We calculate $j_{{\rm s}\parallel}^x$ to leading order in the spin-mixing conductance per unit area, $g_{\uparrow\downarrow}$. To keep the notation simple, we present the calculation for a one-dimensional F$|$N junction. To generalize to the three-dimensional case it is sufficient to replace the magnon density of states $\nu_{\rm m}(\Omega)$ by $\nu_{\rm m}^{3{\rm D}}(\Omega)$.
  
Starting point of our calculation is the Hamiltonian coupling conduction electrons in N and magnons in F,
\begin{equation}
  \hat H = J \hat \psi^{\dagger}_{\uparrow} \hat \psi_{\downarrow} \hat a +
  J^* \hat \psi^{\dagger}_{\downarrow} \hat \psi_{\uparrow} \hat a^{\dagger}.
  \label{eq:Ham}
\end{equation}

\noindent
Here $\hat \psi_{\sigma}$ is the annihilation operator for a conduction electron with spin $\sigma$ at the F$|$N interface, $J$ is the (suitably normalized) interfacial exchange (s-d) interaction strength, and the raising and lowering operators $\hat a^{\dagger}$ and $\hat a$ describe the transverse magnetization amplitude at the F$|$N interface at $x=0$. (They are the Fourier transforms of the second-quantization counterparts of the amplitude $a_{\rm in}(\Omega) + a_{\rm out}(\Omega)$ of the previous section.) The spin current through the F$|$N interface is
\begin{align}
  \label{eq:jsJ}
  \hat j^x_{\rm s \parallel} =&\, 
  i [J \hat \psi^{\dagger}_{\uparrow} \hat \psi_{\downarrow} \hat a \
  - J^* \hat \psi^{\dagger}_{\downarrow} \hat \psi_{\uparrow} \hat a^{\dagger}].
\end{align}

Calculating the expectation value $j^x_{\rm s \parallel}$ to leading order in $J$ using Fermi's Golden rule, one finds
\begin{align}
  j^x_{{\rm s}\parallel} =&\,
  2 \pi |J|^2 \nu^2
  \int\limits_{-\infty}^{\infty} d\varepsilon
  \int\limits_{\omega_0}^{\infty} d\Omega \,
  \nu_{\rm m}(\Omega)
  \\ &\, 
  \mbox{} \times
  \left\{ n_{\uparrow}(\varepsilon) [1 - n_{\downarrow}(\varepsilon-\hbar \Omega)] [1 + f_{T_{\rm F}}(\Omega - \mu_{\rm m}/\hbar)] 
  \right. \nonumber \\ &\, \left. 
  \mbox{}
  - [1-n_{\uparrow}(\varepsilon)] n_{\downarrow}(\varepsilon-\hbar \Omega) f_{T_{\rm F}}(\Omega - \mu_{\rm m}/\hbar)
  \right\} \!, \nonumber
\end{align}

\noindent
where $n_{\sigma}$ is the distribution function of electrons with spin $\sigma$ in N, $\nu$ the electron density of states at the Fermi energy, and $\nu_{\rm m}(\Omega)$ the magnon density of states at the interface. (We assume that the electronic density of states is constant within the energy window of interest.) Taking a Fermi-Dirac distribution with chemical potential $\mu_{\sigma}$ and temperature $T_{\rm N}$ for the electron distribution function $n_{\sigma}$ and performing the integration over the electron energy $\varepsilon$, one obtains
\begin{align}
  \label{eq:Isdc}
  j^x_{{\rm s}\parallel} =&\,
  2 \pi |J|^2 \nu^2
  \int\limits_{\omega_0}^{\infty} d\Omega \,
  \nu_{\rm m}(\Omega)
  (\hbar \Omega - \mu_{\rm s \parallel})
  \nonumber \\ &\,  
  \mbox{} \times
  \left[ f_{T_{\rm N}}(\Omega - \mu_{\rm s \parallel}/\hbar) -
    f_{T_{\rm F}}(\Omega - \mu_{\rm m}/\hbar) \right] \!,
\end{align}

\noindent
where $\mu_{\rm s \parallel} = \mu_{\uparrow} - \mu_{\downarrow}$ and $f_{T}$ is the Planck distribution as before. This result is identical to Eq.\ (\ref{eq:jsperturbative}) if we identify\cite{Bender_2015}
\begin{equation}
  |J|^2 \nu^2 = \frac{\mbox{Re}\, g_{\uparrow\downarrow}}{2 \pi^2 s}.
  \label{eq:identification_w_low-f}
\end{equation}

To obtain the spin current density for an oscillating spin accumulation, we set
\begin{equation}
  \label{eq:muomega}
  \mu_{\sigma}(t) = \bar \mu_{\sigma} +
  \int\limits_{-\infty}^{\infty} d\omega
  \, \delta \mu_{\sigma}(\omega) e^{-i \omega t},\ \
  \mu_{\rm m}(t) = \bar \mu_{\rm m},
\end{equation}
with $\delta \mu_{\sigma}(\omega) = \delta \mu_{\sigma}(-\omega)^*$. Hence, we impose  oscillating chemical potentials $\delta \mu_{\sigma}$ on top of a time-independent background $\bar \mu_{\sigma}$ in N and a time-independent background $\bar \mu_{\rm m}$ in F. We use the method of non-equilibrium Green functions to calculate the expectation value $j^x_{{\rm s}\parallel}$ in the presence of the chemical potentials of Eq.\ (\ref{eq:muomega}). To linear order in $\delta \mu_{\rm s \parallel}(\omega) = \delta \mu_{\uparrow}(\omega) - \delta \mu_{\downarrow}(\omega)$, we find (see App.\ \ref{app:Keldysh} for details)
\begin{equation}
  j^x_{{\rm s}\parallel}(t) = \bar j^x_{{\rm s}\parallel} +
  \int\limits_{-\infty}^{\infty} d\omega \, \delta j^x_{{\rm s}\parallel}(\omega) e^{-i \omega t},
\end{equation}

\noindent
with $\bar j^x_{{\rm s}\parallel}$ equal to the steady-state spin current density of Eq. \eqref{eq:Isdc} with $\mu_{\rm m} = \bar \mu_{\rm m}, \mu_{\rm s \parallel} = \bar \mu_{\rm s \parallel}$ and 
\begin{equation}
  \delta j^x_{{\rm s}\parallel}(\omega) = \frac{g_{{\rm s}\parallel}(\omega)}{4 \pi}
  \delta \mu_{\rm s \parallel}(\omega).
  \label{eq:g_s_parallel_omega}
\end{equation}

\noindent
Here $g_{{\rm s}\parallel}(\omega)$ is the finite-frequency longitudinal spin conductance per unit area,
\begin{align}
  \label{eq:jsperturbativeomega}
  g_{{\rm s}\parallel}(\omega) =&\,
  i \frac{2 \, \mbox{Re}\, g_{\uparrow\downarrow}}{\pi s}  
  \int\limits_{-\infty}^{\infty} d\Omega
  \\ &\, \mbox{} \times
  \left\{
  D(\Omega) \left[
    f_{T_{\rm F}}(\Omega - \bar \mu_{\rm m}/\hbar)
  -  {\cal F}_{\rm N}(\Omega,\omega)
    \right]
  \right. \nonumber \\ &\, \left. \ \ \ \ \mbox{}
  -
  D(\Omega)^* \left[
    f_{T_{\rm F}}(\Omega - \bar \mu_{\rm m}/\hbar)
    - {\cal F}_{\rm N}(\Omega,-\omega)
    \right] \right\} \!,\nonumber 
\end{align}

\noindent
where we defined
\begin{align}
\label{eq:cal_F_N}
  {\cal F}_{\rm N}(\Omega,\omega) =&\,
  \frac{1}{\hbar \omega}
  \left[(\hbar \Omega - \bar \mu_{\rm s \parallel}) f_{T_{\rm N}}(\Omega- \bar \mu_{\rm s\parallel}/\hbar) 
  \right. \\ \nonumber &\, \left. \ \ \ \ \mbox{}
  - (\hbar \Omega-\hbar \omega- \bar \mu_{\rm s \parallel})f_{T_{\rm N}}(\Omega-\omega-\bar \mu_{\rm s \parallel}/\hbar)
    \right] \!,
\end{align}

\noindent
and where
\begin{equation}
  D(\Omega) = \int\limits_{-\infty}^{\infty} d\Omega' \frac{\nu_{\rm m}(\Omega')}{\Omega + i \eta - \Omega'}
\label{eq:magnon_Green_function}
\end{equation}

\noindent
is the (retarded) magnon Green function, with $\eta$ a positive infinitesimal. One verifies that Eq.\ (\ref{eq:jsperturbativeomega}) reproduces the perturbative result in Eq. (\ref{eq:jsperturbative}) for the limit $\omega \to 0$ and that it satisfies the Kramers-Kronig relation
\begin{equation}
  g_{{\rm s}\parallel}(\omega) = \frac{1}{i \pi}
  \int d\omega' \frac{\mbox{Re}\, g_{{\rm s}\parallel}(\omega')}{\omega'-\omega-i \eta}.
\end{equation}

\section{Discussion}
\label{sec:4}

{\em Zero-frequency limit.---}
We evaluate the results of our calculations in Secs.\ \ref{sec:2} and \ref{sec:3} for the paradigmatic spintronic material combination YIG$|$Pt. Longitudinal spin transport through the F$|$N interface is expected to play an important role for the ferrimagnetic insulator YIG, since at room temperature the longitudinal spin conductance $g_{{\rm s}\parallel}$ is comparable to the (transverse) spin-mixing conductance $g_{\uparrow\downarrow}$ for this material. (This leads, {\em e.g.}, to a prediction of a remarkable frequency dependence of the spin-Hall magnetoresistance for this material combination.\cite{Reiss_2021}) To facilitate a comparison with the literature, we use the same material parameters as Cornelissen {\em et al.} in Ref. \onlinecite{Cornelissen_2016} (if applicable). We summarize the material parameters in Tab. \ref{tab:estimates_parameters}. 

\begin{table}
\centering
\begin{tabular*}{\columnwidth}{@{\extracolsep{\fill}}lll}
\hline
\hline
material & experimental parameters & ref. \\
\hline
YIG & $\omega_0/2 \pi = 7.96 \cdot \rm{10^{9} \, Hz}$ & \cite{Hahn_2013}\\
& $a_{\rm m} = 1.2 \cdot 10^{-9}\rm{m}$ & \cite{Cornelissen_2016} \\
& $\Dex = 8.0 \cdot 10^{-6} \, \rm{m}^2 \, \rm{s}^{-1}$ & \cite{Cornelissen_2016}\\ 
& $s = 5.3 \cdot 10^{27} \rm{m}^{-3}$& \cite{Cornelissen_2016}\\
\hline
YIG$|$Pt & $(e^2/h) \textrm{Re} \, g_{\uparrow \downarrow} = 1.6 \cdot 10^{14} \, \Omega^{-1} \rm{m}^{-2} $ & \cite{Hahn_2013,Qiu_2013}\\
 & $(e^2/h) \textrm{Im} \, g_{\uparrow \downarrow} = 0.08 \cdot 10^{14} \, \Omega^{-1} \rm{m}^{-2} $ & \\
\hline
\hline
\end{tabular*}
\caption{Typical values for the relevant material parameters of YIG and YIG$|$Pt interfaces considered in this article. The last column states the references used for our estimates. The spin density $s = S/a_{\rm m}^3$, where $S = 10$ is the magnitude of the spin in each magnetic unit cell with lattice constant $a_{\rm m}$. The frequency of acoustic magnons at the zone boundary is $\Omega_{\rm max}/2 \pi \approx (12\Dex/a_{\rm m}^2)/2 \pi \approx 1.0 \times 10^{13}\, {\rm Hz}$. The imaginary part of the spin-mixing conductance is found from the estimate $\textrm{Im} \, g_{\uparrow \downarrow}/\textrm{Re} \, g_{\uparrow \downarrow} \approx 0.05$, see Refs.\ \onlinecite{Jia_2011,Althammer_2013}.}
\label{tab:estimates_parameters}
\end{table}

Our non-perturbative calculation of the longitudinal spin conductance uses the magnon dispersion of the Landau-Lifshitz equation \eqref{eq:llg}. This is a good approximation at long wavelengths, for which the magnon dispersion is quadratic as in Eq.\ (\ref{eq:kx}). The use of the quadratic approximation to the magnon dispersion is justified if $k_{\rm B} T \ll \hbar \Omega_{\rm max}$, where $\Omega_{\rm max}$ is the frequency of acoustic magnons at the zone boundary,
\begin{equation}
  \Omega_{\rm max} \approx \omega_0 + \frac{12 \Dex}{a_{\rm m}^2},
  \label{eq:Omegamax}
\end{equation}
with $a_{\rm m}$ the size the of the magnetic unit cell. For YIG, one has $\Omega_{\rm max}/2 \pi \approx 
10^{13}\, {\rm Hz}$,\cite{Cherepanov_1993,Barker_2016} so that the condition $k_{\rm B} T \ll \hbar \Omega_{\rm max}$ is only weakly obeyed at room temperature.

\begin{figure}
\centering
\includegraphics[width=0.45\textwidth]{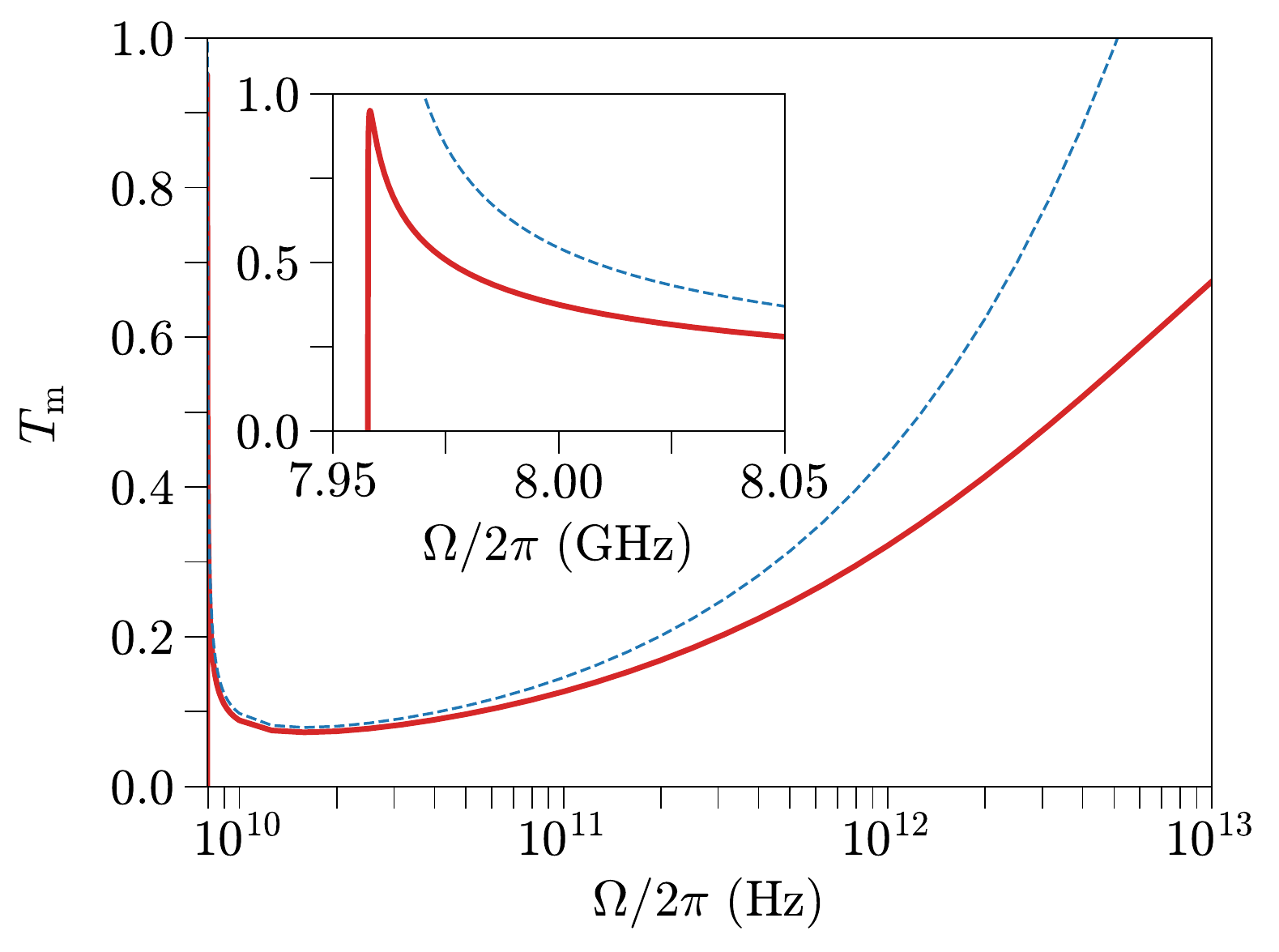}
\caption{Mode-averaged magnon transmission coefficient $T_{\rm m} (\Omega)$ at a YIG$|$Pt-interface. The solid red curve shows the non-perturbative result of Eq.\ (\ref{eq:Tm}) and the blue dashed curve the weak-coupling approximation $T_{\rm m}^{({\rm p})}(\Omega)$ of Eq.\ (\ref{eq:trans_prob_approx}). Both curves are based on the quadratic approximation to the magnon dispersion, which breaks down for magnon frequencies $\Omega \approx \Omega_{\rm max}$, which is the frequency of acoustic magnons at the zone boundary. Parameter values are taken from Tab. \ref{tab:estimates_parameters}.\label{fig:magnon_trans_prob}}
\end{figure}

The result \eqref{eq:Tm} for the mode-averaged transmission coefficient $T_{\rm m} (\Omega)$, which is the probability that a magnon is annihilated at the F$|$N interface and excites a spinful excitation of the conduction electrons in N, is shown in Fig. \ref{fig:magnon_trans_prob} for $\mu_{{\rm s}\parallel} = 0$.
At the lowest magnon frequency $\omega_0$, the magnon wave vector $\vk = 0$ ({\em i.e.}, $k_x = k_{\perp} = 0$) and thus the reflection coefficient $\rho (\omega_0, k_{\perp}) = -1$, so that $T_{\rm m}(\omega_0) = 0$. 
However, upon increasing $\Omega$ above $\omega_0$, $|\rho (\Omega, k_{\perp})|$ first very quickly drops to approximately $0$ and then reaches a maximum; correspondingly, $T_{\rm m} (\Omega)$ first features a maximum and then reaches a minimum upon increasing the magnon frequency $\Omega$ above $\omega_0$. The maximum is at a frequency $(\Omega - \omega_0)/\omega_0 \approx \omega_0 |g_{\uparrow\downarrow}|^2/(4 \pi s)^2 \Dex \ll 1$; the minimum is at $\Omega \approx 2 \omega_0$. Upon further increasing the frequency, $T_{\rm m}(\Omega)$ increases monotonously with $\Omega$. In this frequency range, a good approximation for $T_{\rm m} (\Omega)$ is obtained by expanding $|\tau(\Omega, k_{\perp})|^2$ to first order in $g_{\uparrow\downarrow}$, which gives
\begin{align}
  T_{\rm m}^{({\rm p})}(\Omega) =
  \frac{8 \pi \mbox{Re}\, g_{\uparrow\downarrow}}{s} \frac{(\Omega - \mu_{{\rm s}\parallel}/\hbar) \nu_{\rm m}^{\rm 3D} (\Omega)}{k_x(\Omega)^2}
\label{eq:trans_prob_approx}
\end{align}

\noindent
as shown by the blue dashed curve in Fig. \ref{fig:magnon_trans_prob}.
The perturbative approximation for $T_{\rm m}(\Omega)$ remains valid for $\Omega \ll (4 \pi s)^2 \Dex/|g_{\uparrow\downarrow}|^2$, a condition that is obeyed as long as $\Omega \ll \Omega_{\rm max}$. (The condition $\Omega \ll (4 \pi s)^2 \Dex/|g_{\uparrow\downarrow}|^2$ becomes equal to the condition $\Omega \ll \Omega_{\rm max}$ if one uses the Sharvin approximation for the spin-mixing conductance\cite{Brataas_2000,Zwierzycki_2005} and takes the Fermi wavelength of electrons in N to be of the same order of magnitude as the size $a_{\rm m}$ of the magnetic unit cell, so that $|g_{\uparrow\downarrow}| \approx 1/a_{\rm m}^2$.)

\begin{figure}
\centering
\includegraphics[width=0.45\textwidth]{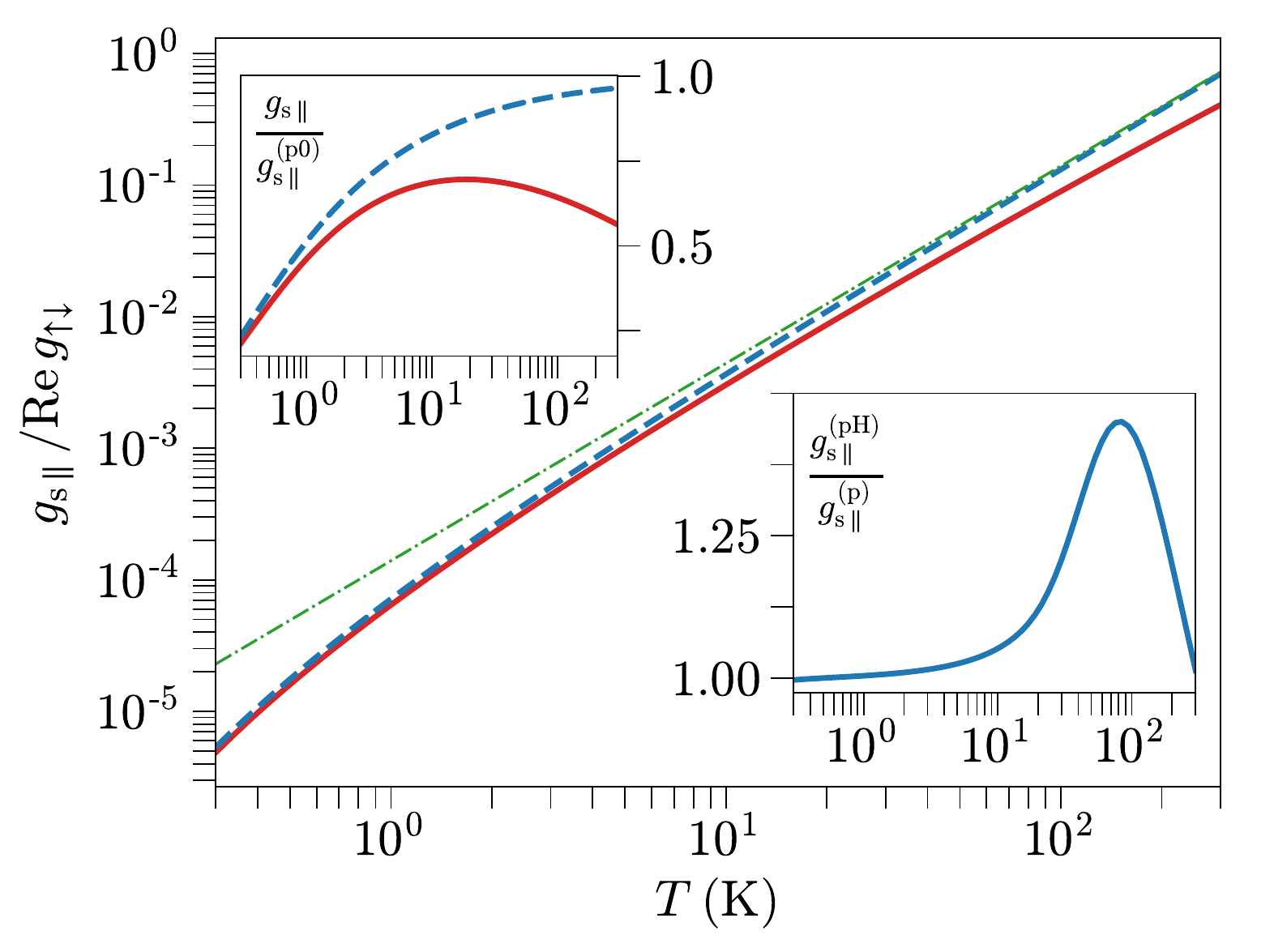}
\caption{Zero-frequency longitudinal spin conductance per unit area, $g_{{\rm s}\parallel}$, at a YIG$|$Pt-interface as function of the temperature $T = T_{\rm N} = T_{\rm F}$ for $\mu = \mu_{{\rm s}\parallel} = \mu_{\rm m} = 0$. The red solid curve shows the non-perturbative result $g_{{\rm s}\parallel}/\text{Re } g_{\uparrow\downarrow}$ of Eq.\ \eqref{eq:gs1}, the blue dashed curve the perturbative result $g_{{\rm s}\parallel}^{({\rm p})}/\text{Re } g_{\uparrow\downarrow}$ of Eq.\ \eqref{eq:gs2}, and the thin green dot-dashed curve the approximation $g_{{\rm s}\parallel}^{({\rm p0})}/\text{Re } g_{\uparrow\downarrow}$ of Eq.\ \eqref{eq:gsimpleT}. The upper left inset shows the ratios $g_{{\rm s}\parallel}/g_{{\rm s}\parallel}^{({\rm p}0)}$ (red solid curve) and $g_{{\rm s}\parallel}^{({\rm p})}/g_{{\rm s}\parallel}^{({\rm p}0)}$ (blue dashed curve). The lower right inset shows the ratio $g_{{\rm s}\parallel}^{({\rm pH})}/g_{{\rm s}\parallel}^{({\rm p})}$, where $g_{{\rm s}\parallel}^{({\rm pH})}$ is the result of Eq. \eqref{eq:gs2} for the magnon density of states obtained from a Heisenberg model, see Eq.\ (\ref{eq:dispersion_Heisenberg}), and $g_{{\rm s}\parallel}^{({\rm p})}$ that of Eq. \eqref{eq:gs2} for the quadratic approximation of the magnon dispersion.
Parameter values are taken from Tab. \ref{tab:estimates_parameters}.
\label{fig:low-freq_gsparallel_ratio_lin_response}}
\end{figure}

Now we are ready to discuss the differential longitudinal spin conductance per unit area
\begin{equation}
  g_{{\rm s}\parallel} = 4 \pi \frac{\partial j_{{\rm s}\parallel}^x}{\partial \mu_{{\rm s}\parallel}}.
\end{equation}
{}From Eq.\ (\ref{eq:jszero3d}) we find for $T = T_{\rm N} = T_{\rm F}$ and $\mu = \mu_{{\rm s}\parallel} = \mu_{\rm m}$, that
\begin{align}
  g_{{\rm s}\parallel} =&\, \frac{1}{2 \pi} \int\limits_{\omega_0}^{\infty} d\Omega \, k_x(\Omega)^2 T_{\rm m}(\Omega) \left(- \frac{\partial f_{T}(\Omega - \mu/\hbar )}{\partial \Omega} \right) \!.
  \label{eq:gs1}
\end{align}
In the perturbative limit of small $g_{\uparrow\downarrow}$ this result simplifies to
\begin{align}
  g_{{\rm s}\parallel}^{({\rm p})} =&\,
  \frac{4 \mbox{Re}\, g_{\uparrow\downarrow}}{s} 
  \int\limits_{\omega_0}^{\infty} d\Omega \, \nu^{\rm 3D}_{\rm m}(\Omega)
  (\Omega - \mu/\hbar)
  \nonumber \\ &\, \mbox{} \times
  \left(- \frac{\partial f_{T}(\Omega - \mu/\hbar)}{\partial \Omega} \right) \!.
  \label{eq:gs2}
\end{align}
The perturbative result for the ratio $g_{{\rm s}\parallel}/\mbox{Re}\, g_{\uparrow\downarrow}$ depends on the magnetic properties of bulk YIG only and not on the choice of the normal metal N or the transparency of the interface, whereas the non-perturbative result shows a (quantitative, but not qualitative) dependence on the interface properties. The results of Eqs. (\ref{eq:gs1}) and (\ref{eq:gs2}) are shown in Fig. \ref{fig:low-freq_gsparallel_ratio_lin_response} as functions of temperature $T$ for the material parameters of a YIG$|$Pt interface, see Tab.\ \ref{tab:estimates_parameters}. (We assume no temperature dependence of the spin density $s$ and the spin stiffness $\Dex$.)
The green dashed straight line in Fig.\ \ref{fig:low-freq_gsparallel_ratio_lin_response} is the perturbative result with the additional approximation $\hbar \omega_0 \ll k_{\rm B} T$, which gives\cite{Cornelissen_2016}
\begin{equation} 
  g_{{\rm s}\parallel}^{({\rm p}0)} = c \frac{\mbox{Re}\, g_{\uparrow\downarrow}}{s}
  \left[
    \left( \frac{k_{\rm B} T_{\rm F}}{\pi \hbar \Dex} \right)^{3/2}
  + \frac{1}{2} \left( \frac{k_{\rm B} T_{\rm N}}{\pi \hbar \Dex} \right)^{3/2} \right]
    \ \
    \label{eq:gsimpleT}
\end{equation}
with $c = (1/2) \zeta(3/2) \approx 1.31$. The difference between the perturbative and non-perturbative results increases with temperature and reaches a factor $\approx 1.7$ at room temperature, whereby the non-perturbative result for $g_{{\rm s}\parallel}$ is always below the small-$g_{\uparrow\downarrow}$ approximation, see Fig.\ \ref{fig:low-freq_gsparallel_ratio_lin_response} (upper left inset).

Since the perturbative finite-frequency expression for the longitudinal spin conductance, discussed below, can not be evaluated using a magnon density of states $\nu_{\rm m}(\Omega)$ of a continuum magnon model, we compare the zero-frequency longitudinal spin conductance for a quadratic magnon dispersion (as is used in the main panel of Fig.\ \ref{fig:low-freq_gsparallel_ratio_lin_response}) with that for a magnon dispersion of a Heisenberg model on a simple cubic lattice (see Eq.\ (\ref{eq:dispersion_Heisenberg}) below). This comparison is shown in the lower right inset of Fig.\  \ref{fig:low-freq_gsparallel_ratio_lin_response}.
Whereas the difference between the two cases is small for low temperatures and near room temperature, the Heisenberg model leads to a longitudinal spin conductance that is up to a factor $\approx 1.45$ larger than that of the quadratic approximation at intermediate temperatures. This is consistent with the absence of van Hove peaks in the magnon density of states in the quadratic approximation.

In principle, the differential longitudinal spin conductance per unit area, $g_{{\rm s}\parallel}$, also depends on the chemical potentials $\mu_{{\rm s}\parallel}$ and $\mu_{\rm m}$. Such dependence governs the interfacial spin current beyond linear order in $\mu_{{\rm s}\parallel}- \mu_{\rm m}$. 
Because the driving potentials $\mu_{{\rm s}\parallel}$ and $\mu_{\rm m}$ must remain below $\hbar \omega_0$ --- otherwise the magnon system is unstable ---, the range of admissible values for $\mu_{{\rm s}\parallel}$ and $\mu_{\rm m}$ remains well below $k_{\rm B} T$ at most temperatures, so that appreciable nonlinear effects can be found only for extremely low temperatures $T \lesssim 1\, {\rm K}$. At those low temperatures thermal magnons are as good as absent, so that the longitudinal spin conductance is negligibly small in comparison to the transverse spin conductance. For a further discussion we refer to the discussion of nonlinear effects in the context of the finite-frequency longitudinal spin conductance below.

{\em Finite-frequency longitudinal spin transport.---} For a discussion of the finite-frequency longitudinal spin conductance per unit area, $g_{{\rm s}\parallel}(\omega)$, the quadratic approximation of the magnon dispersion is not sufficient even at temperatures $k_{\rm B} T \ll \hbar \Omega_{\rm max}$. The reason is that at finite frequencies, $g_{{\rm s}\parallel}(\omega)$ acquires a finite imaginary part, which depends on the full magnon spectrum. (The real part of $g_{{\rm s}\parallel}(\omega)$, which describes the dissipative response, can still be calculated within the quadratic approximation.) 
For temperatures of the order of room temperature and below and for frequencies $\omega \lesssim \Omega_{\rm max}$ 
it is sufficient to consider the lowest-lying magnon band and neglect higher magnon bands 
in YIG.\cite{Barker_2016} The lowest magnon band can be described effectively by a Heisenberg model of spins on a simple cubic lattice with nearest-neighbor interactions.\cite{Cherepanov_1993,Plant_1983} The resulting dispersion relation is given by 
\begin{align}
\Omega(\vk) = \omega_0 + \frac{2 D_{\rm ex}}{a_{\rm m}^2}  \sum\limits_{\alpha = x,y,z} (1-\cos (k_{\alpha} a_{\rm m})),
\label{eq:dispersion_Heisenberg}
\end{align}
with maximal magnon frequency $\Omega_{\rm max}$, given in Eq.\ (\ref{eq:Omegamax}), and  agrees with the quadratic approximation for $\Omega \ll \Omega_{\rm max}$. 
The finite magnon bandwidth
regularizes the integrations for the imaginary part of $g_{{\rm s}\parallel}(\omega)$.
In the numerical evaluations of the real and imaginary parts of $g_{{\rm s}\parallel}(\omega)$ that are discussed below we therefore use the magnon density of states corresponding to the dispersion in Eq. (\ref{eq:dispersion_Heisenberg}). We verified that as long as $\omega$, $k_{\rm B} T/\hbar \ll \Omega_{\rm max}$ the results for real and imaginary parts of $g_{{\rm s}\parallel}(\omega)$ depend only weakly on the precise form of the magnon density of states at frequencies $\omega \gg k_{\rm B} T/\hbar$.

\begin{figure}
\centering
\includegraphics[width=0.45\textwidth]{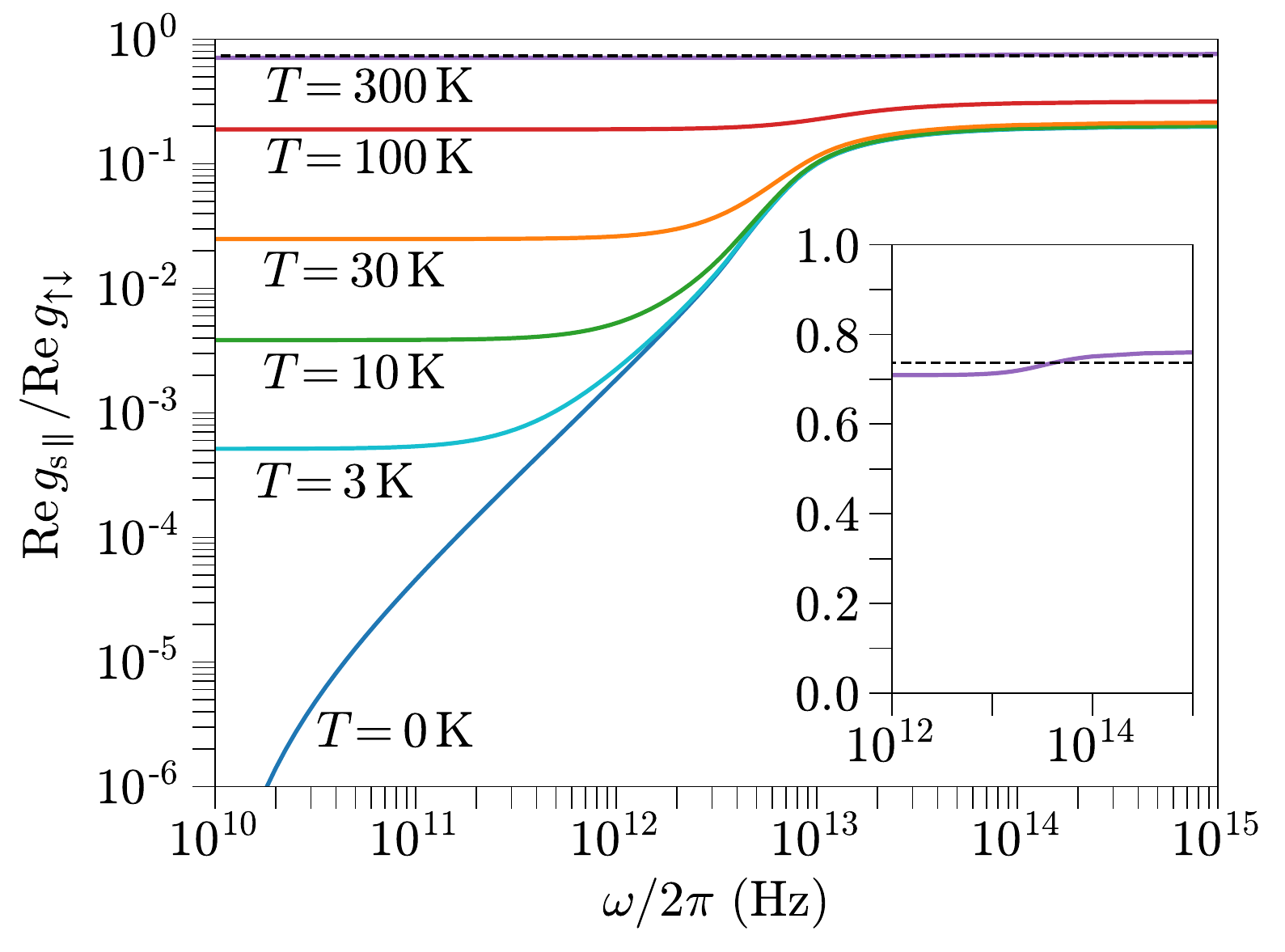}
\caption{Real part $\mbox{Re}\, g_{{\rm s}\parallel}$ of the finite-frequency longitudinal spin conductance of an F$|$N interface in the perturbative regime of weak coupling as a function of the driving frequency $\omega/2 \pi$ for various temperatures $T = T_{\rm N} = T_{\rm F}$ (solid colored lines). Material parameters are taken for an F$|$N interface with F=YIG and N an arbitrary normal metal, see Tab.\ \ref{tab:estimates_parameters}. The black dashed curve shows the result of the Rayleigh-Jeans approximation, see Eq. \eqref{eq:Rayleigh-Jeans}. The inset displays the same curves. The time-independent background magnon chemical potential and spin accumulation have been set to zero, $\bar \mu_{\rm m} = \bar \mu_{\rm s \parallel} = 0$.
\label{fig:ratio_real_part_FFSC} 
}
\end{figure}

\begin{figure}
\centering
\includegraphics[width=0.45\textwidth]{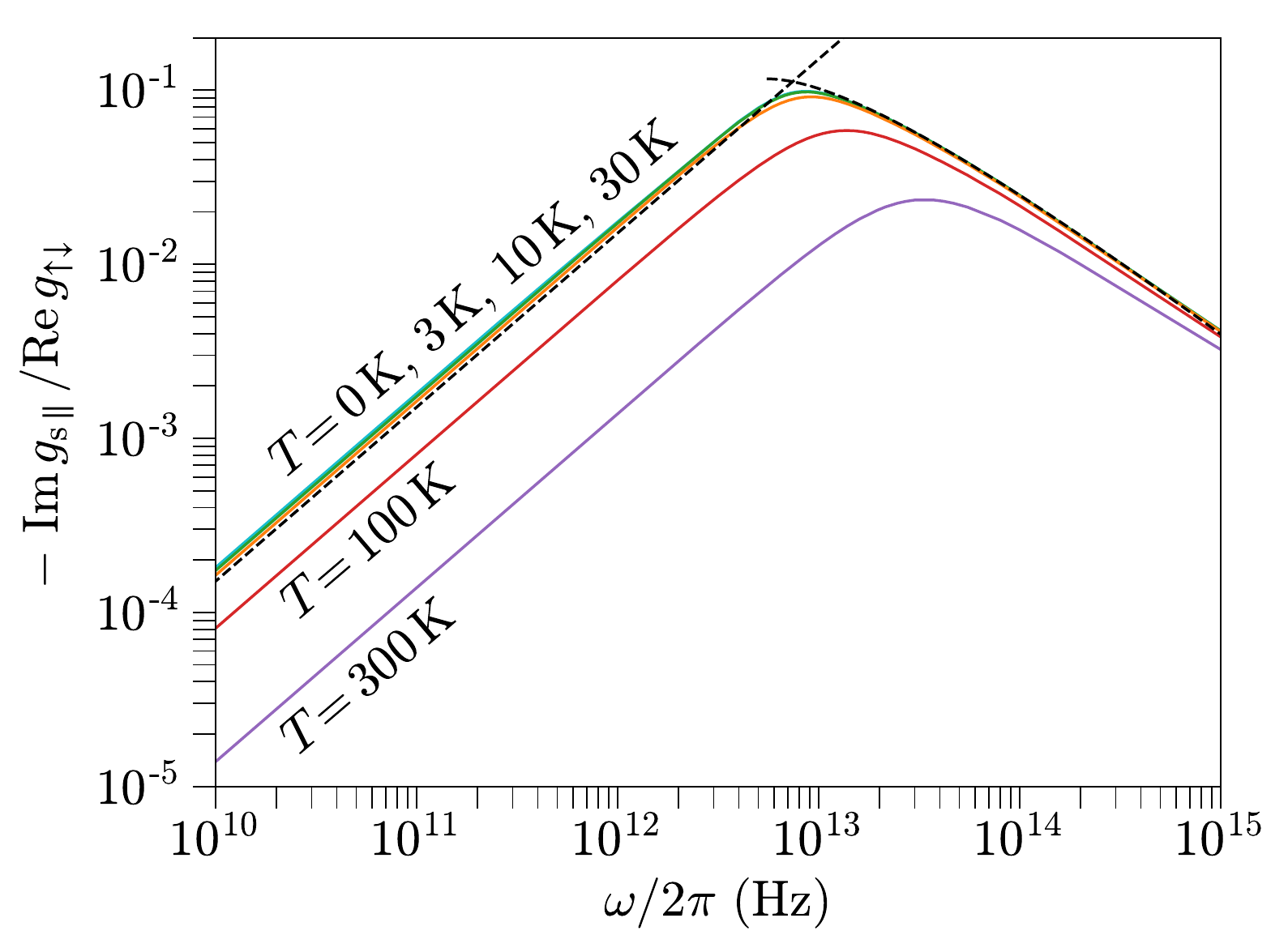}
\caption{Same as Fig.\  \ref{fig:ratio_real_part_FFSC}, but for the imaginary part $\mbox{Im}\, g_{{\rm s}\parallel}$ of the finite-frequency spin conductance. The black dashed lines show the limiting behavior for small and large $\omega$ according to Eqs. \eqref{eq:limit_small_omega} and \eqref{eq:limit_large_omega} for $T_{\rm N} \to 0$. 
\label{fig:ratio_imaginary_part_FFSC}
}
\end{figure}

Figures \ref{fig:ratio_real_part_FFSC} and \ref{fig:ratio_imaginary_part_FFSC} show the real and imaginary parts of the finite-frequency spin conductance $g_{{\rm s}\parallel}$ at an F$|$N interface with F=YIG as function of the driving frequency $\omega$ and for different temperatures $T = T_{\rm N} = T_{\rm F}$ and $\bar \mu_{{\rm s}\parallel} = \bar \mu_{{\rm m}} = 0$. In the perturbative regime, the ratio $g_{{\rm s}\parallel}/\mbox{Re}\, g_{\uparrow\downarrow}$ is independent of the choice of the normal metal N or the quality of the F$|$N interface.

For driving frequencies $\omega \ll k_{\rm B} T/\hbar$, the real part $\mbox{Re}\, g_{{\rm s}\parallel}$ approaches the zero-frequency limit discussed above.
(Note that there may be small deviations between the zero-frequency limit obtained from the quadratic approximation of the magnon dispersion and from the magnon dispersion of Eq.\ (\ref{eq:dispersion_Heisenberg}), see Fig.\ \ref{fig:low-freq_gsparallel_ratio_lin_response}, lower right inset.)
For $T = 300\,{\rm K}$, the real part $\mbox{Re}\, g_{{\rm s}\parallel}$ does not show an appreciable frequency dependence. At this temperature, the Planck distribution $f_{T_{\rm N}}$ may be well approximated by the Rayleigh-Jeans distribution 
\begin{align}
f_{T_{\rm N}}(\Omega) = \frac{k_{\rm B} T_{\rm N}}{\hbar \Omega}. \label{eq:Rayleigh-Jeans}
\end{align}

\noindent
In this limit one finds that ${\cal F}_{\rm N}(\Omega,\omega) = 0$ in Eq.\ (\ref{eq:jsperturbativeomega}), so that $g_{{\rm s}\parallel}(\omega)$ is independent of  frequency $\omega$, temperature $T_{\rm N}$, and background spin accumulation $\bar \mu_{{\rm s}\parallel}$ in N. At lower temperatures, $\mbox{Re}\, g_{{\rm s}\parallel}$ shows an increase with frequency for $\omega \gtrsim k_{\rm B} T_{\rm N}/\hbar$, followed by a saturation at $\omega \approx \Omega_{\rm max}$. One may obtain an analytical expression for $\mbox{Re}\, g_{{\rm s}\parallel}(\omega)$ in the limit $\omega \gg k_{\rm B} T_{\rm N}/\hbar$ (setting $\bar \mu_{{\rm s}\parallel} = \bar \mu_{\rm m} = 0$):
\begin{align}
  \label{eq:gs3}
  \mbox{Re}\, g_{{\rm s}\parallel}(\omega) \approx&\,
  \frac{2 \mbox{Re}\, g_{\uparrow\downarrow}}{s}
  \left[ 2 \vphantom{\int_{0}^{M^M_M}}
  \int\limits_{\omega_0}^{\infty} d\Omega \, \nu_{\rm m}(\Omega) f_{T_{\rm F}}(\Omega - \bar \mu_{\rm m})
  \right. \nonumber \\ &\, \left. \mbox{} +
  \int\limits_{\omega_0}^{\omega + \bar \mu_{{\rm s}\parallel}/\hbar} d\Omega \, \nu_{\rm m}(\Omega)
  \frac{\omega - \Omega - \bar \mu_{{\rm s}\parallel}}{\omega} \right] \!.
\end{align}
(To keep the notation simple, we drop the superscript ``(p)'' because all finite-frequency spin conductances are obtained in the perturbative limit of small $g_{\uparrow\downarrow}$.)
The first line in Eq.\ (\ref{eq:gs3}) is a frequency-independent offset which depends on the temperature $T_{\rm F}$ and magnon chemical potential $\bar \mu_{\rm m}$ of the ferro-/ferrimagnetic insulator only. Using the quadratic approximation for the magnon dispersion and assuming $\omega_0 \ll k_{\rm B} T_{\rm F}/\hbar$, this term is found to be equal to the first term in Eq.\ (\ref{eq:gsimpleT}).
For $\omega_0$, $k_{\rm B} T_{\rm N}/\hbar \ll \omega \ll \Omega_{\rm max}$ we may also use the quadratic approximation for the magnon dispersion in the second term and find
\begin{align}
  \mbox{Re}\, g_{{\rm s}\parallel}(\omega) \approx&\,
  \frac{\mbox{Re}\, g_{\uparrow\downarrow}}{s}
  \left[
  c  \left( \frac{k_{\rm B} T_{\rm F}}{\pi \hbar \Dex} \right)^{3/2}
  + \frac{8}{15}
    \left(\frac{\omega}{\Dex} \right)^{3/2}
  \right] \!,
  \label{eq:gsimpleomega}
\end{align}
where $c = (1/2) \zeta(3/2) \approx 1.31$ as below Eq.\ (\ref{eq:gsimpleT}).
In the limit $\omega \gg \Omega_{\rm max}$ (but still $k_{\rm B} T \ll \hbar \Omega_{\rm max}$) we find similarly
\begin{align}
  \mbox{Re}\, g_{{\rm s}\parallel}(\omega) \approx&\, \frac{\mbox{Re}\, g_{\uparrow\downarrow}}{s} \left[
   c \left( \frac{k_{\rm B} T_{\rm F}}{\pi \hbar \Dex} \right)^{3/2}
  +  \frac{2}{a_{\rm m}^3} \right] \!.
  \label{eq:gsimpleomegamax}
\end{align}

\noindent
where $a_{\rm m}$ is the lattice constant of the magnetic unit cell. (Note that, up to a numerical factor of order unity in the second term, Eq.\ (\ref{eq:gsimpleomegamax}) is what one obtains when $k_{\rm B} T_{\rm N}/\hbar$ in Eq.\ (\ref{eq:gsimpleT}) is replaced by $\Omega_{\rm max}$.)

With respect to the high-frequency limit $\omega \gtrsim \Omega_{\rm max}$ and/or the high-temperature limit $k_{\rm B} T \gtrsim \hbar \Omega_{\rm max}$, it should be kept in mind that our calculation only considers the contribution from the lowest-lying magnon band. For such high frequencies, other magnon bands are likely to contribute to $g_{{\rm s}\parallel}(\omega)$ as well and such a contribution is not included in our theory. Hence, Eq.\ (\ref{eq:gsimpleomegamax}) and analogously Eq. (\ref{eq:limit_large_omega}) for $\mbox{Im}\, g_{{\rm s}\parallel}(\omega)$ discussed below should be interpreted as the contribution of the lowest-lying magnon band to the longitudinal spin conductance only. 

The imaginary part $\mbox{Im}\, g_{{\rm s}\parallel}(\omega)$ increases linearly with $\omega$ for small frequencies, reaches a maximum at max($\Omega_{\rm max}, k_{\rm B} T/\hbar$), and decreases with $\omega$ in the high-frequency limit, see Fig.\ \ref{fig:ratio_imaginary_part_FFSC}. The linear increase with $\omega$ for frequencies $\omega \lesssim \Omega_{\rm max}$ is given by the expression
\begin{align}
  \mbox{Im}\, g_{{\rm s}\parallel}(\omega) \approx&\,
  - \omega \frac{2 \mbox{Re}\, g_{\uparrow\downarrow}}{\pi s}
  \int
  d\Omega \frac{\nu_{\rm m}(\Omega + \bar \mu_{{\rm s}\parallel}/\hbar)}{\Omega}
  h_{T_{\rm N}}(\Omega),
  \label{eq:limit_small_omega}
\end{align}
with
\begin{align}
  h_T(\Omega) =&\, \int\limits_{- \infty}^{\infty} d\Omega' \frac{\Omega}{\Omega - \Omega'}
  \frac{\partial^2}{\partial \Omega'^2} \left[\Omega' f_{T}(\Omega')\right] \!.
\end{align}
This function behaves as $h_T(\Omega) \to 1$ for $\Omega \gg k_{\rm B} T/\hbar$ 
and $h_T(\Omega) \to 0$ for $\Omega \ll k_{\rm B} T/\hbar$. Hence, effectively only frequencies $\Omega \gtrsim k_{\rm B} T/\hbar$ contribute to the integration in Eq.\ (\ref{eq:limit_small_omega}), which explains the decreasing slope of $- \mbox{Im}\, g_{{\rm s}\parallel}(\omega)$ vs.\ $\omega$ --- {\em i.e.}, the intercept with the vertical axis in Fig.\ \ref{fig:ratio_imaginary_part_FFSC} --- with increasing temperature $T$. The decay of $\mbox{Im}\, g_{{\rm s}\parallel}(\omega)$ in the limit of large frequencies $\omega \gg \Omega_{\rm max}$ is described by
\begin{align}
   \label{eq:limit_large_omega}
  \mbox{Im}\, g_{{\rm s}\parallel}(\omega) \approx&\,
   - \frac{1}{\omega} \frac{4 \mbox{Re}\, g_{\uparrow\downarrow}}{\pi s} \int d\Omega \, \nu_{\rm m}(\Omega + \bar \mu_{{\rm s}\parallel}\hbar) \Omega
  \nonumber \\  &\, \ \ \ \ \mbox{} \times
   \left[ 
   1 + \ln \frac{\omega}{\Omega} + h'_{T_{\rm N}}(\Omega) \right] \!,
\end{align}  

\noindent
with
\begin{equation}
  h'_{T}(\Omega) = \int\limits_{- \infty}^{\infty} d\Omega' \frac{\Omega'}{\Omega(\Omega'-\Omega)} \left[ f_{T}(\Omega') + \Theta (- \Omega') \right],
\end{equation}

\noindent
where $\Theta (\Omega')$ is the Heaviside step function. The temperature-dependent term proportional to $h'_{T_{\rm N}}$ is sub-leading for $\omega \gg \Omega_{\rm max}$, so that $\mbox{Im}\, g_{{\rm s}\parallel}(\omega)$ becomes effectively temperature-independent for sufficiently high frequency $\omega$, as seen in Fig.\ \ref{fig:ratio_imaginary_part_FFSC}.



\begin{figure}
\centering
\includegraphics[width=0.45\textwidth]{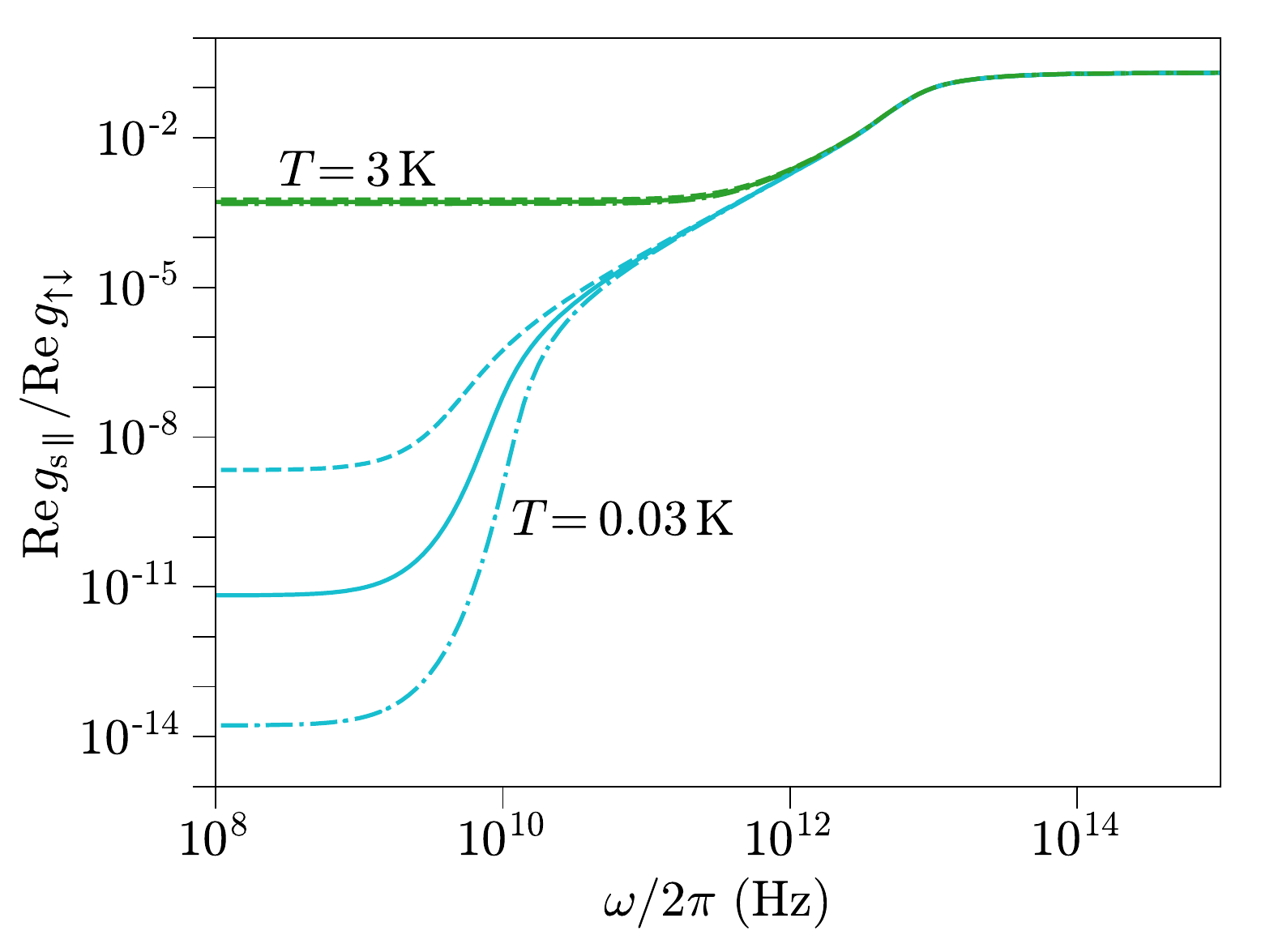}
\caption{Real part $\mbox{Re}\, g_{{\rm s}\parallel}$ of the finite-frequency spin conductance of an F$|$N interface in the weak-coupling regime as function of the driving frequency for two values of the temperature $T = T_{\rm N} = T_{\rm F}$. Material parameters are taken for an F$|$N interface with F=YIG and N is an arbitrary normal metal, see Tab.\ \ref{tab:estimates_parameters}. Curves are shown for three combinations of the time-independent background potentials $\bar \mu_{\rm m}$ and $\bar \mu_{{\rm s}\parallel}$: The solid colored curves correspond to $\bar \mu_{\rm m} = \bar \mu_{\rm s \parallel} = 0$; the dashed curves correspond to $\bar \mu_{\rm m} = \bar \mu_{\rm s \parallel} = 0.5 \hbar \omega_0$, and the dot-dashed ones to $\bar \mu_{\rm m} = \bar \mu_{\rm s \parallel} = - 0.5 \hbar \omega_0$.
\label{fig:ratio_real_part_FFSC_small_mu}
}
\end{figure}

The role of the time-independent background magnon chemical potential $\bar \mu_{{\rm m}}$ and spin accumulation $\bar \mu_{{\rm s}\parallel}$ is addressed in Fig.\ \ref{fig:ratio_real_part_FFSC_small_mu}. The figure shows $\mbox{Re}\, g_{{\rm s}\parallel}(\omega)$ as function of $\omega$, as in Fig. \ref{fig:ratio_real_part_FFSC}, but for different values of $\bar \mu_{\rm m}$ and $\bar \mu_{\rm s \parallel}$, while satisfying the bound $\bar \mu_{\rm m}$, $\bar \mu_{{\rm s}\parallel} < \hbar \omega_0$. As the magnon chemical potential and spin accumulation appear in Eq. \eqref{eq:jsperturbativeomega} only in the combinations $\bar \mu_{\rm m}/k_{\rm B} T_{\rm F}$ and $\bar \mu_{\rm s \parallel}/k_{\rm B} T_{\rm N}$ and since $\hbar \omega_0$ is much smaller than $k_{\rm B} T$ for most temperatures considered, we only show results for $T = T_{\rm N} = T_{\rm F} = 0.03\,$K and $T = T_{\rm N} = T_{\rm F} = 3\,$K. 
As can be seen in Fig.\ \ref{fig:ratio_real_part_FFSC_small_mu}, the dependence of $\mbox{Re}\, g_{{\rm s}\parallel}(\omega)$ on $\bar \mu_{{\rm m}}$ and $\bar \mu_{{\rm s}\parallel}$ disappears, when $\hbar \omega_0 \ll k_{\rm B} T$ (as for $T = 3$K in Fig. \ref{fig:ratio_real_part_FFSC_small_mu}) or when $\hbar \omega$ becomes large in comparison to $\bar \mu_{\rm m}$ and $\bar \mu_{{\rm s}\parallel}$. The imaginary part of $g_{{\rm s}\parallel}(\omega)$ does not show any appreciable dependence on $\bar \mu_{{\rm s}\parallel}$ in the full parameter range considered (not shown) and is independent of $\bar \mu_{\rm m}$.

\begin{figure}
\centering
\includegraphics[width=0.45\textwidth]{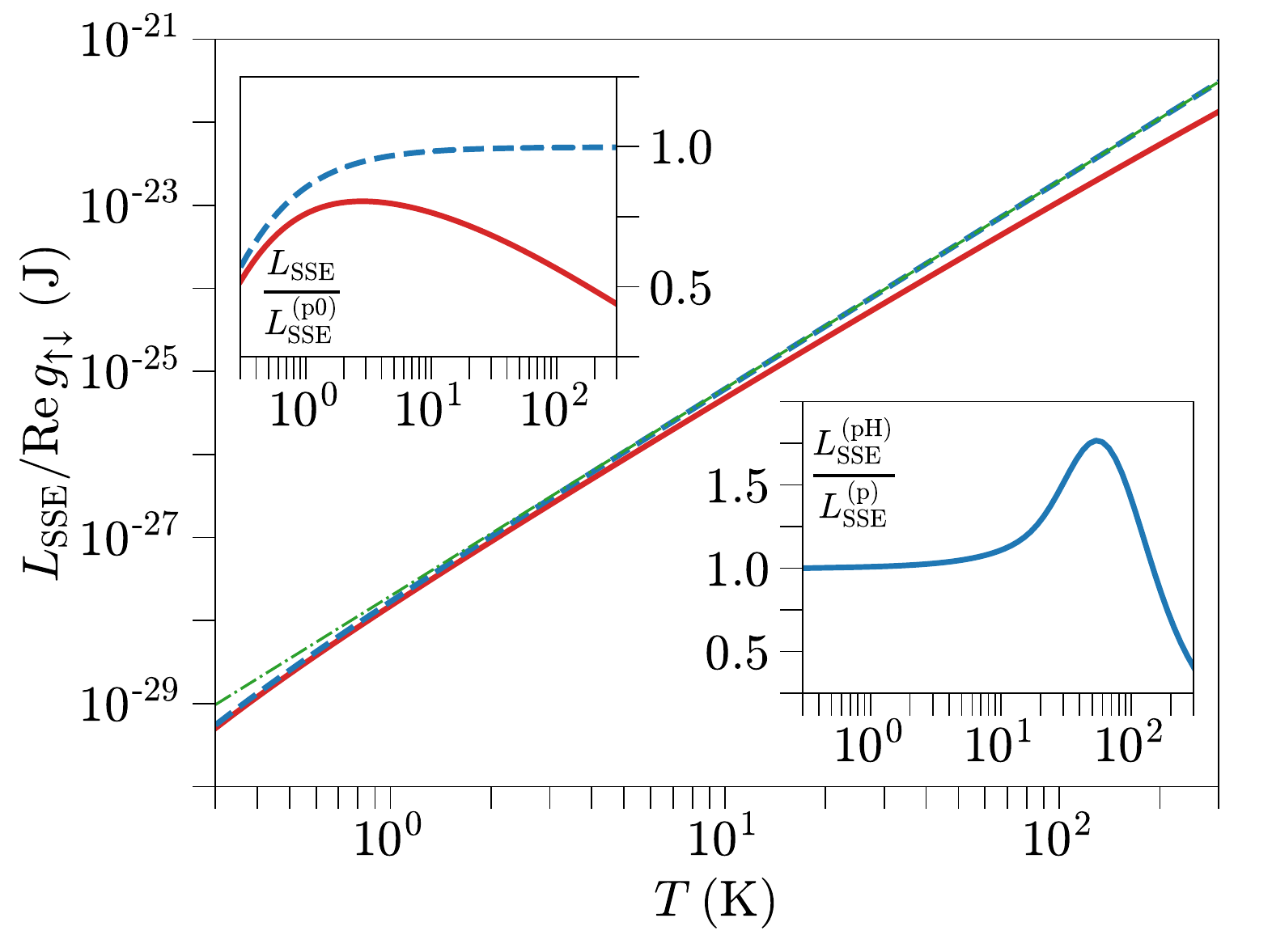}
\caption{Spin-Seebeck coefficient $L_{\rm SSE}$ at a YIG$|$Pt-interface in the linear-response regime as function of the temperature $T = \bar T_{\rm N} = \bar T_{\rm F}$. All three curves in the main panel are obtained using the parabolic approximation of the magnon dispersion. The red solid curve shows $L_{\rm SSE}$ according to the non-perturbative theory, see Eq. \eqref{eq:L_SSE_preciser}, the blue dashed curve the perturbative result $L_{\rm SSE}^{({\rm p})}$ of Eq. \eqref{eq:L_SSE_approx}, and the thin green dot-dashed curve includes the approximation $L_{\rm SSE}^{({\rm p}0)}$ of Eq.\ (\ref{eq:L_SSE_approx2}). The upper left inset shows the ratios $L_{\rm SSE}/L_{\rm SSE}^{({\rm p}0)}$ (red solid curve) and $L_{\rm SSE}^{({\rm p})}/L_{\rm SSE}^{({\rm p}0)}$ (blue dashed curve). 
 The lower right inset shows the ratio $L_{\rm SSE}^{({\rm pH})}/L_{\rm SSE}^{({\rm p})}$, where $L_{\rm SSE}^{({\rm pH})}$ is the result of Eq. \eqref{eq:L_SSE_approx} for the magnon density of states obtained from a Heisenberg model, see Eq.\ (\ref{eq:dispersion_Heisenberg}), and $L_{\rm SSE}^{({\rm p})}$ that of Eq. \eqref{eq:L_SSE_approx} for the quadratic approximation of the magnon dispersion.
Parameter values are taken from Tab. \ref{tab:estimates_parameters}. 
\label{fig:L_SSE_ratio_lin_response}}
\end{figure}

{\em Spin-Seebeck coefficient.---} Our non-perturbative calculation of the longitudinal spin current through the F$|$N interface also describes the longitudinal spin current in response to a temperature difference $\delta T$ across the interface. We set $T_{\rm F} = T + \delta T$, $T_{\rm N} = T$, $\mu_{\rm m}= \mu_{{\rm s}\parallel}$, and expand $j^x_{{\rm s}\parallel}$ in Eq.\ \eqref{eq:jszero3d} to linear order in $\delta T$, resulting in
\begin{align}
  j^x_{{\rm s}\parallel} = \frac{L_{\rm SSE}}{T} \delta T
\end{align}
with the spin-Seebeck coefficient $L_{\rm SSE}$\cite{Schmidt_2021}
\begin{align}
L_{\rm SSE} = &\, \frac{\hbar}{2 (2 \pi)^2} \int\limits_{\omega_0}^{\infty} d\Omega\, k_x(\Omega)^2
  T_{\rm m} (\Omega) (\Omega - \mu_{{\rm s}\parallel}/\hbar)
\nonumber\\
& \mbox{} \times  \left( - \frac{\partial f_T(\Omega - \mu_{{\rm s}\parallel}/\hbar)}{\partial \Omega} \right) \!.
\label{eq:L_SSE_preciser}
\end{align}
In the weak-coupling limit of Eq. \eqref{eq:jsperturbative}, one recovers the spin-Seebeck coefficient obtained by Cornelissen {\em et al.},\cite{Cornelissen_2016}
\begin{align}
  L_{\rm SSE}^{({\rm p})} = &\, \frac{\hbar \text{Re} \, g_{\uparrow \downarrow}}{\pi s} \int d\Omega \, \nu_{\rm m}^{\rm 3D}(\Omega) ( \Omega - \mu_{{\rm s}\parallel}/\hbar)^2
  \nonumber\\&\, \mbox{} \times
  \left( - \frac{\partial f_T(\Omega - \mu_{{\rm s}\parallel}/\hbar)}{\partial \Omega} \right) \!.
\label{eq:L_SSE_approx}
\end{align}
In the limit $\omega_0 \ll k_{\rm B} T/\hbar$ the frequency integration may be performed and one finds\cite{Cornelissen_2016}
\begin{equation}
    L_{\rm SSE}^{({\rm p0})} = c' \frac{\text{Re} \, g_{\uparrow \downarrow} k_{\rm B} T}{\pi s} \left( \frac{k_{\rm B} T}{\pi \hbar \Dex} \right)^{3/2},
  \label{eq:L_SSE_approx2}
\end{equation}
with $c' = 15\,\zeta(5/2)/32 \approx 0.63$. 
All three expressions are evaluated in Fig.\ \ref{fig:L_SSE_ratio_lin_response} as a function of $T$ for material parameters of a YIG$|$Pt interface. Like in the case of the longitudinal spin conductance, we observe that there are small quantitative differences between the non-perturbative and perturbative results. These differences are small at low temperatures, but the perturbative weak-coupling result deviates from the non-perturbative one at higher temperatures, the difference reaching a factor $\approx 2.3$ at room temperature, see the upper left inset of Fig.\ \ref{fig:L_SSE_ratio_lin_response}. 


\section{Conclusions and outlook}
\label{sec:5}

The spin angular momentum current from a normal metal N into a ferro-/ferrimagnetic insulator F in general has a component collinear with the magnetization, which is carried by thermal magnons in F. In this article, we presented two calculations of the longitudinal interfacial spin conductance: At zero frequency, but for arbitrary transparency of the interface, and at finite frequencies, but to leading order in the interface transparency. 
In general, one expects the longitudinal interfacial spin conductance to acquire a dependence on the driving frequency $\omega$, when $\omega$ exceeds $k_{\rm B} T/\hbar$. In the case of typical parameters for the material combination YIG$|$Pt and at room temperature, we find that the resulting frequency dependence of the interfacial spin conductance is rather weak, not more than a factor $\approx 1.1$ between the low- and high-frequency limits. Also, we find that (at zero frequency) the difference between the spin conductance in a 
non-perturbative treatment of the coupling across the F$|$N interface and the perturbative result to leading order in the spin-mixing conductance $g_{\uparrow \downarrow}$ is not more than a factor $\approx 1.7$ at room temperature, despite the fact that $g_{\uparrow \downarrow}$ of a good YIG$|$Pt interface (see Tab. \ref{tab:estimates_parameters}) is only slightly below the Sharvin limit $(e^2/h) g_{\uparrow \downarrow} = \pi e^2/h\lambda_e^2 \approx 6.8 \cdot 10^{14}\,  \Omega^{-1}$m$^{-2}$,\cite{Zwierzycki_2005} where $\lambda_{e}$ is the Fermi wavelength of Pt.\cite{Ketterson_1969,Mueller_1971,Fradin_1975} In that sense, for F$|$N interfaces involving the ferrimagnetic insulator YIG, our two calculations may seen as a confirmation of the existing low-frequency weak-coupling theory.\cite{Zhang_2012,Bender_2015,Cornelissen_2016} A similar conclusion applies to the interfacial spin-Seebeck coefficient, for which we compared the existing weak-coupling zero-frequency theory\cite{Xiao_2010,Cornelissen_2016,Schmidt_2018} with a calculation non-perturbative in the interface transparency.

Of course, one may turn the question around and ask, under which experimental conditions or for which material combinations a frequency dependence of the interfacial longitudinal spin conductance or a deviation from the perturbative weak-coupling approximation will become significant. To see an appreciable frequency dependence of $g_{{\rm s}\parallel}$, it is necessary that the temperature is significantly below the maximum energy $\hbar \Omega_{\rm max}$ of acoustic magnons. For YIG, this means that the temperature must be well below room temperature. Our numerical estimates based on material parameters for YIG indicate that $g_{{\rm s}\parallel}(\omega)$ may increase by a factor $\gtrsim 10$ between low- and high-frequency regimes if $T \lesssim 30\,{\rm K}$ and that the effect can be larger at lower temperatures, whereas the frequency dependence of $g_{{\rm s}\parallel}(\omega)$ is small for $T \gtrsim 100\,{\rm K}$.

An experimental technique to measure these effects is the spin-Hall magnetoresistance, which depends on the competition of longitudinal and transversal spin transport across the F$|$N interface. Measurements of the spin-Hall magnetoresistance up to the lower GHz range\cite{Lotze_2014}
have already been performed. Since the longitudinal and transversal interfacial spin conductances are of comparable magnitude in the high-frequency limit, one may thus expect a visible frequency dependence of the spin-Hall magnetoresistance effect for frequencies in the THz range, if the temperature is low enough that not all magnon modes are thermally excited. (This effect is additional to a frequency dependence of the spin-Hall magnetoresistance in the GHz range predicted in Ref.\ \onlinecite{Reiss_2021}.) However, since the spin-Hall magnetoresistance effect involves the difference of two contributions of comparable magnitude, a more precise material-specific modeling is required to reach a firm prediction.

Another experimental platform in which the longitudinal interfacial spin conductance plays a role is that of non-local magnonic spin transport.\cite{Zhang_2012,Goennenwein_2015,Schlitz_2021} In this case, the interfacial spin conductance directly determines the coupling between the magnon system in a ferro-/ferrimagnetic insulator and the electrical currents in adjacent normal-metal contacts used to excite and detect the magnon currents. Our predictions directly translate to a frequency dependence of the electron-to-magnon and magnon-to-electron conversion in such experiments. Furthermore, the difference between the weak-coupling and strong-coupling predictions may quantitatively affect estimates of the spin-mixing conductance based on a measurement of the longitudinal spin conductance or the spin-Seebeck coefficient.\cite{Hahn_2013,Hahn_2013_2,Althammer_2013,Nakayama_2013,Wang_2014,Vlietstra_2013,Qiu_2013}.

We predict that the longitudinal spin conductance depends on the temperatures $T_{\rm F}$ and $T_{\rm N}$ of the ferro-/ferrimagnetic insulator and the normal metal in different ways, see, {\em e.g.}, Eqs.\ (\ref{eq:gsimpleT}) and (\ref{eq:gsimpleomegamax}). Whereas the longitudinal spin current in F is carried by thermal magnons if F and N are close to equilibrium, the longitudinal spin conductance does not vanish if $T_{\rm F} = 0$, as long as $T_{\rm N}$ is non-zero. In this case, the spin current is carried by magnons in F excited by spin-flip scattering of thermally excited electrons at the F$|$N interface. Apart from the difficulty that a large temperature difference between F and N is difficult to realize experimentally, a large temperature difference across an F$|$N interface also leads to a large steady-state spin current via the interfacial spin-Seebeck effect. However, this DC spin current can be easily distinguished experimentally from the AC signal, which is caused by time-dependent driving of the spin accumulation in N.

At the interface between a normal metal and a ferro-/ferrimagnetic metal, there are two contributions to the longitudinal spin current: A contribution from conduction electrons in the ferro-/ferrimagnetic metal and a magnonic contribution. The results derived in this article also apply to the magnonic contribution at such an interface. However, at metal-metal interfaces, the magnonic contribution to the spin current is typically much smaller than the electronic contribution so that the frequency and temperature dependence of the magnonic contribution is a sub-leading effect at such interfaces.

We close with two remarks on possible further extensions of our work.
An important limitation of our theory is the restriction to the lowest magnon band. On the one hand, this limitation enters into our non-perturbative calculation for low frequencies, because the calculation relies on the continuum limit of the Landau-Lifshitz-Gilbert equation. On the other hand, this limitation enters both calculations, because the boundary condition at the F$|$N interface implicitly assumes that the coupling between electronic degrees of freedom in N and the magnonic degrees of freedom in F at the interface is local. For acoustic magnons at the zone boundary and for higher magnon bands, electrons in N reflecting off the ferro-/ferrimagnetic insulator F penetrate F sufficiently deep such that they are influenced by the non-uniformity of $\vm$, violating the assumption of a local coupling between magnonic and electronic degrees of freedom. The first problem can be partially addressed by replacing the quadratic magnon dispersion by the dispersion of a Heisenberg model on a simple cubic lattice, as we have done in Sec.\ \ref{sec:4}, but this replacement does not account for the non-uniformity of the magnetization near the interface. A rough estimate for the frequency at which the non-uniformity becomes relevant is the maximum frequency $\Omega_{\rm max}$ of acoustic magnons, where for YIG $\Omega_{\rm max}/2 \pi \, \approx 10^{13}\,{\rm Hz}$. 
It is an open task for the future to extend our theory to appropriate couplings between electron spins and short-wavelength magnons, optical magnons, and antiferromagnons in antiferromagnets and ferrimagnets.

Our finite-frequency calculations assume that it is only the electronic distribution in the normal metal N that is driven out of equilibrium. Experiments exciting directly the phonons of an insulating magnet F such as YIG, \textit{e.g.}, by an ultrashort THz laser pulse, might also create a time-dependent magnon chemical potential in F on ultrafast time scales.\cite{Cornelissen_2016} Time-dependent magnon chemical potentials may also appear in ultrafast versions of non-local magnon transport experiments or in the ultrafast spin-Hall magnetoresistance effect with an ultrathin magnetic insulator F.\cite{Reiss_2021} Investigating the ultrafast response to a change of magnon chemical potential is another interesting avenue for future research.

{\em Acknowledgements.---}
We thank T. Kampfrath, T. S. Seifert, U. Atxitia, F. Jakobs, and S. M. Rouzegar for stimulating discussions. This work was funded by the German Research Foundation (DFG) via the collaborative research center SFB-TRR 227 ``Ultrafast Spin Dynamics" (project B03).

\appendix

\section{Transformation to a rotating frame}
\label{app:a}

Here we consider the transformation to a reference system for the spin degree of freedom that rotates with angular frequency $\bar \vomega(t) = \bar \omega(t) \ve_{\parallel}$. We discuss how the longitudinal spin accumulation $\mu_{{\rm s}\parallel}$ in N, the magnetization amplitude $m_{\perp}$, and the stochastic transverse spin current $j_{{\rm s}\perp}$ transform upon passing to the rotating frame. We restrict the discussion to linear response in $\mu_{{\rm s}\parallel}$ and $\bar \vomega$.
We use a prime to denote creation and annihilation operators and observables in the rotating reference system.

We first consider the transformation to a frame rotating at constant angular velocity $\bar \vomega = \bar \omega \ve_{\parallel}$. The transformation relation for the electron annihilation operators in N is
\begin{equation}
  \hat \psi'(t) = e^{i \bar \vomegascr \cdot \vsigmascr t/2} \hat \psi(t),
\end{equation}
where $\psi(t)$ is a two-component column spinor for the wavefunction of the conduction electrons. Solving for the annihilation operator $c(\varepsilon)$ in energy representation, we have
\begin{align}
  \hat \psi'(\varepsilon) =&\,
  \hat \psi(\varepsilon + \hbar \bar \vomega \cdot \vsigma/2)
\end{align}
and, similarly,
\begin{align}
  \hat \psi'(\varepsilon)^{\dagger} =&\,
  \hat \psi(\varepsilon + \hbar \bar \vomega \cdot \vsigma/2)^{\dagger}.
\end{align}
It follows that the distribution function $f'(\varepsilon)$ in the rotating frame is
\begin{align}
  f_{T_{\rm N}}'(\varepsilon) = f_{T_{\rm N}}(\varepsilon + \hbar \bar \vomega \cdot \vsigma/2),
\end{align}
where $f_{T_{\rm N}}$ is the distribution function in the original reference frame. We thus conclude that, in linear response, upon transforming to a rotating frame the spin accumulation changes as
\begin{equation}
  \vmu_{\rm s}' = \vmu_{\rm s} - \hbar \bar \vomega.
\end{equation}

\section{Weak-coupling spin current at finite frequency}
\label{app:Keldysh}

We first discuss the expression (\ref{eq:jsJ}) for the longitudinal spin current through the F$|$N interface. From the Heisenberg equation of motion, the spin current into the magnetic insulator is 
\begin{align}
  \hat j^x_{{\rm s}\parallel}(t) =&\, - \frac{i}{2} [\hat H,\hat N_{\uparrow} - \hat N_{\downarrow}],
\end{align}
where $\hat H$ is the Hamiltonian and $\hat N_{\sigma}$ is the number of conduction electrons with spin $\sigma$, $\sigma = \uparrow$, $\downarrow$. The only contribution to $\hat H$ that does not commute with $\hat N_{\sigma}$ is the term (\ref{eq:Ham}) describing the coupling via the F$|$N interface. Inserting Eq\ (\ref{eq:Ham}) into the above expression gives Eq.\ (\ref{eq:jsJ}) of the main text.

We next turn to the calculation of the expectation value $j^x_{{\rm s}\parallel}(t)$ of the interfacial longitudinal spin current. Calculating $j^x_{{\rm s}\parallel}(t)$ to leading order in perturbation theory in $J$ gives
\begin{align}
  \label{eq:Is1}
  j^x_{{\rm s}\parallel}(t) =&\,
  i \frac{|J|^2}{\hbar} \int_{\rm c} dt'
  \left[ G_{\uparrow}(t',t) G_{\downarrow}(t,t') D(t,t')
  \right. \nonumber \\ &\, \left. \ \ \ \ \mbox{}
    -
    G_{\uparrow}(t,t') G_{\downarrow}(t',t) D(t',t) \right] \!,
\end{align}
where $t'$ is integrated along the Keldysh contour ({\em i.e.}, forward and backward integrations along the real time axis),
\begin{equation}
  G_{\sigma}(t',t) = -i \langle T_{\rm c} \hat \psi_{\sigma}(t') \hat \psi_{\sigma}^{\dagger}(t) \rangle
\end{equation}
is the contour-ordered Green function for the conduction electrons, evaluated at the interface, and
\begin{equation}
  D(t',t) = -i \langle T_{\rm c} \hat a(t') \hat a^{\dagger}(t) \rangle
\end{equation}
is the contour-ordered magnon Green function, again evaluated at the interface.
Equation (\ref{eq:Is1}) may be written as
\begin{widetext}
\begin{align}
  \label{eq:Is2}
  j^x_{{\rm s}\parallel}(t)  =&\,
  i \frac{|J|^2}{\hbar} \int\limits_{-\infty}^{\infty} dt'
  \left[(G^{\rm R}_{\uparrow}(t',t) + G^{<}_{\uparrow}(t',t))
  (G^{\rm R}_{\downarrow}(t,t') + G^{<}_{\downarrow}(t,t'))  
  (D^{\rm R}(t,t') + D^{<}(t,t'))
  \right. \nonumber \\ &\, \ \ \ \ \left. \mbox{}  
  - (G^{\rm R}_{\uparrow}(t,t') + G^{<}_{\uparrow}(t,t'))
  (G^{\rm R}_{\downarrow}(t',t) + G^{<}_{\downarrow}(t',t))  
  (D^{\rm R}(t',t) + D^{<}(t',t)) 
  \right. \nonumber \\ &\, \ \ \ \ \left. \mbox{}  
  - G_{\uparrow}^{>}(t',t) G_{\downarrow}^{<}(t,t') D^{<}(t,t')
  + G_{\uparrow}^{<}(t,t') G_{\downarrow}^{>}(t',t) D^{>}(t',t) \right] \!.
\end{align}
\end{widetext}
In this expression, the integration variable $t'$ is a time, not a contour time.

We first evaluate Eq.\ (\ref{eq:Is2}) for the case that the three subsystems --- conduction electrons with spin up, conduction electrons with spin down, and magnons --- are separately in equilibrium at chemical potentials $\mu_{\sigma}$ and $\mu_{\rm m}$ and temperatures $T_{\sigma}$ and $T_{\rm m}$, respectively. In this case, all Green functions depend on the time difference $t-t'$ only. Changing to the integration variable $t'-t$ for the first term and third term and $t-t'$ for the second and fourth term in Eq.\ (\ref{eq:Is2}), one finds that the first and third terms in Eq.\ (\ref{eq:Is2}) cancel, whereas the second and fourth terms give, after Fourier transform,
\begin{align}
  j^x_{{\rm s}\parallel} =&\,
  - i \frac{|J|^2}{\hbar} \int \frac{d\varepsilon}{2 \pi}
  \int \frac{d\Omega}{2 \pi}
  \left[G_{\uparrow}^{>}(\varepsilon) G_{\downarrow}^{<}(\varepsilon-\Omega)
    D^{<}(\Omega)
    \right. \nonumber \\ &\, \left. \ \ \ \ \mbox{}
    - G_{\uparrow}^{<}(\varepsilon) G_{\downarrow}^{>}(\varepsilon-\Omega)
    D^{>}(\Omega)
    \right] \!.
\end{align}

\noindent
According to the fluctuation-dissipation theorem, one has
\begin{align}
  G_{\sigma}^>(\varepsilon) =&\, -2 \pi i \hbar (1-n_{\sigma}(\varepsilon)) \nu_{\sigma},\nonumber \\
  G_{\sigma}^<(\varepsilon) =&\, 2 \pi i \hbar n_{\sigma}(\varepsilon) \nu_{\sigma}.
  \label{eq:fluctuation_dissipation_1}
\end{align}

\noindent
Similarly, for the magnon Green function, one has
\begin{align}
  D^>(\Omega) =&\, -2 \pi i (f_{T_{\rm F}}(\Omega - \mu_{\rm m}/\hbar)+1) \nu_{\rm m}(\Omega),\nonumber \\
  D^<(\Omega) =&\, -2 \pi i f_{T_{\rm F}}(\Omega - \mu_{\rm m}/\hbar) \nu_{\rm m}(\Omega).
  \label{eq:fluctuation_dissipation_2} 
\end{align}

\noindent
Hence, we find that the spin current is
\begin{align}
  j^x_{{\rm s}\parallel} =&\,
  2 \pi |J|^2 \nu_{\uparrow} \nu_{\downarrow}
  \int d\varepsilon
  \int d\Omega \,
  \nu_{\rm m}(\Omega)
  \\ &\, \ \ \ \ \mbox{} \times
  \left[ n_{\uparrow}(\varepsilon) (1 - n_{\downarrow}(\varepsilon-\hbar \Omega)) (1 + f_{T_{\rm F}}(\Omega) - \mu_{\rm m}/\hbar)
  \right. \nonumber \\ &\, \left. \ \ \ \ \mbox{}
  - (1-n_{\uparrow}(\varepsilon))n_{\downarrow}(\varepsilon-\hbar \Omega) f_{T_{\rm F}}(\Omega - \mu_{\rm m}/\hbar)
    \right] \!.\nonumber 
\end{align}

\noindent
Setting $T_{\uparrow} = T_{\downarrow} = T_{\rm N}$ and performing the integration over $\varepsilon$, one reproduces Eqs. (29) and (30) of the main text, which was derived from Fermi's Golden Rule.

We now consider an additional oscillating component of the chemical potential as in Eq.\ (\ref{eq:muomega}) of the main text. In the presence of the oscillating chemical potential the electron Green function $G(t,t')$ reads
\begin{widetext}
\begin{align}
  G_{\sigma}(t,t') =&\, G_{\sigma 0}(t,t') e^{-i \int_{t'}^{t} d\tau \delta \mu_{\sigma}(\tau')/\hbar}
  \nonumber \\ =&\,
  G_{\sigma 0}(t,t')
  \left\{ 1
  + \int d\omega \frac{\delta \mu_{\sigma}(\omega)}{\hbar \omega}
  e^{-i \omega t} [1 - e^{i \omega(t-t')}]
  \right\}
  + \ldots,
\end{align}
for the greater and lesser Green functions, where, in the second line, the subscript ``0'' indicates the equilibrium Green function and the dots indicate terms of higher order in $\delta \mu_{\sigma}(\omega)$. Similarly, one has
\begin{align}
  G_{\sigma}(t',t) = &\, 
  G_{\sigma 0}(t',t)
  \left\{ 1
  - \int d\omega \frac{\delta \mu_{\sigma}(\omega)}{\hbar \omega}
  e^{-i \omega t} [1 - e^{i \omega(t-t')}]
  \right\}
  + \ldots.
\end{align}

To find the spin current, we find it advantageous to cast the first two terms of Eq.\ (\ref{eq:Is2}) into a different form, making repeated use of the identities $G^{\rm R} + G^{<} = G^{\rm A} + G^{>}$ and $D^{\rm R} + D^{<} = D^{\rm A} + D^{>}$,
\begin{align}
  \label{eq:Is3}
  \delta j^x_{{\rm s}\parallel}(t)  =&\,
  i \frac{|J|^2}{\hbar} \int dt'
  \left[(G^{\rm A}_{\uparrow}(t',t) + G^{>}_{\uparrow}(t',t))
  (G^{\rm R}_{\downarrow}(t,t') + G^{<}_{\downarrow}(t,t'))  
  (D^{\rm R}(t,t') + D^{<}(t,t'))
  \right. \nonumber \\ &\, \ \ \ \ \left. \mbox{}  
  - (G^{\rm R}_{\uparrow}(t,t') + G^{<}_{\uparrow}(t,t'))
  (G^{\rm A}_{\downarrow}(t',t) + G^{>}_{\downarrow}(t',t))  
  (D^{\rm A}(t',t) + D^{>}(t',t)) 
  \right. \nonumber \\ &\, \ \ \ \ \left. \mbox{}  
  - G_{\uparrow}^{>}(t',t) G_{\downarrow}^{<}(t,t') D^{<}(t,t')
  + G_{\uparrow}^{<}(t,t') G_{\downarrow}^{>}(t',t) D^{>}(t',t) \right] \!.
\end{align}
For the linear-response correction to the spin current, we then obtain
\begin{align}
  \delta j^x_{{\rm s}\parallel}(\omega) =&\,
  i \frac{|J|^2}{\hbar^2 \omega} \int dt'
  [e^{i \omega(t-t')}-1]
  \nonumber \\ &\, \mbox{} \times
  \left\{
  [ G^{<}_{\uparrow 0}(t'-t)
    G^{\rm R}_{\downarrow}(t-t') \delta \mu_{\uparrow}(\omega)
    - G^{\rm A}_{\uparrow}(t'-t)
  G^{<}_{\downarrow 0}(t-t') \delta \mu_{\downarrow}(\omega)]
  [D^{\rm R}(t-t') + D^{<}(t-t')]
  \right. \nonumber \\ &\, \left. \ \ \mbox{}
  + [ G^{\rm A}_{\downarrow}(t'-t)
    G^{<}_{\uparrow 0}(t-t') \delta \mu_{\uparrow}(\omega)
    - G^{<}_{\downarrow 0}(t'-t)
    G^{\rm R}_{\uparrow}(t-t') \delta \mu_{\downarrow}(\omega)]
  [D^{\rm A}(t'-t) + D^{>}(t'-t)]  
  \right. \nonumber \\ &\, \left. \ \ \mbox{}
  +  G^{>}_{\uparrow 0}(t'-t) G^{<}_{\downarrow 0}(t-t')
  (\delta \mu_{\uparrow}(\omega) - \delta \mu_{\downarrow}(\omega))
    D^{\rm R}(t-t')
  \right. \nonumber \\ &\, \left. \ \ \mbox{}
  + G^{<}_{\uparrow 0}(t-t') G^{>}_{\downarrow 0}(t'-t)
  (\delta \mu_{\uparrow}(\omega) - \delta \mu_{\downarrow}(\omega))
    D^{\rm A}(t'-t)
  \right\} \!.
\end{align}
Writing the Green functions in terms of their Fourier representations, we write this as
\begin{align}
  \delta j^x_{{\rm s}\parallel}(\omega) =&\,
  i \frac{|J|^2}{\hbar \omega}
  \int \frac{d\varepsilon}{2 \pi}
  \int \frac{d\Omega}{2 \pi}
  \nonumber \\ &\, \mbox{} \times \left\{
    \left[
    [G^<_{\uparrow 0}(\varepsilon - \omega) - G^<_{\uparrow 0}(\varepsilon)]
    G^{\rm R}_{\downarrow}(\varepsilon-\Omega) \delta \mu_{\uparrow}(\omega)
    - G^{\rm A}(\varepsilon + \Omega)
    [G^<_{\downarrow 0}(\varepsilon + \omega) - G^<_{\downarrow 0}(\varepsilon)]
    \delta \mu_{\downarrow}(\omega) \right] [D^{\rm R}(\Omega) + D^<(\Omega)]
    \right. \nonumber \\ &\, \left. \ \ \mbox{}
    + \left[ G^<_{\uparrow 0}(\varepsilon + \omega) - G^<_{\uparrow 0}(\varepsilon)]
    G^{\rm A}_{\downarrow}(\varepsilon - \Omega) \delta \mu_{\uparrow}(\omega)
    - G^{\rm R}(\varepsilon + \Omega)
    [G^<_{\downarrow 0}(\varepsilon - \omega) - G^<_{\downarrow 0}(\varepsilon)]
    \delta \mu_{\downarrow}(\omega) \right] [D^{\rm A}(\Omega) + D^>(\Omega)]   
    \right. \nonumber \\ &\, \left. \ \ \mbox{}
    +
    [G^>_{\uparrow 0}(\varepsilon - \omega) - G^>_{\uparrow 0}(\varepsilon)]
    G^<_{\downarrow 0}(\varepsilon - \Omega)
    (\delta \mu_{\uparrow}(\omega) - \delta \mu_{\downarrow}(\omega))
    D^{\rm R}(\Omega)
    \right. \nonumber \\ &\, \left. \ \ \mbox{}
    +
    [G^<_{\uparrow 0}(\varepsilon + \omega) - G^<_{\uparrow 0}(\varepsilon)]
    G^>_{\downarrow 0}(\varepsilon - \Omega)  
    (\delta \mu_{\uparrow}(\omega) - \delta \mu_{\downarrow}(\omega))
    D^{\rm A}(\Omega)
    \right\} \!.
\end{align}
Again we use the fluctuation-dissipation theorem, see Eqs. \eqref{eq:fluctuation_dissipation_1} and \eqref{eq:fluctuation_dissipation_2}. For the electrons we assume that the spectral density is independent of energy and we set $G^{\rm R}_{\sigma}(\varepsilon) = - G^{\rm A}_{\sigma}(\varepsilon) = -i \pi \hbar \nu_{\sigma}$. For the magnons we use that
\begin{align}
  D^{\rm R}(\Omega) + D^<(\Omega) =&\, D^{\rm A}(\Omega) + D^>(\Omega)
  \nonumber \\ =&\, D^{\rm R}(\Omega)(f_{T_{\rm F}}(\Omega - \bar \mu_{\rm m}/\hbar)+1) - D^{\rm A}(\Omega) f_{T_{\rm F}}(\Omega - \bar \mu_{\rm m}/\hbar).
\end{align}
We then find 
\begin{align}
  \delta j^x_{{\rm s}\parallel}(\omega) =&\,
  i \frac{|J|^2}{2 \hbar \omega}
  \int d\varepsilon
  \int d\Omega \, \nu_{\uparrow} \nu_{\downarrow}
  \left\{
  \left[ [n_{\uparrow}(\varepsilon-\hbar\omega)-n_{\uparrow}(\varepsilon+\hbar\omega)]
  \delta \mu_{\uparrow}(\omega)
  - [n_{\downarrow}(\varepsilon-\hbar\omega)-n_{\downarrow}(\varepsilon+\hbar\omega)]
  \delta \mu_{\downarrow}(\omega) \right]
  \right. \nonumber \\ &\, \left.  \ \ \ \ \mbox{} \times
  [D^{\rm R}(\Omega)(f_{T_{\rm F}}(\Omega - \bar \mu_{\rm m}/\hbar)+1) - D^{\rm A}(\Omega) f_{T_{\rm F}}(\Omega - \bar \mu_{\rm m}/\hbar)]
  - 2 (\delta \mu_{\uparrow}(\omega) - \delta \mu_{\downarrow}(\omega))
  \right. \nonumber \\ &\, \left. \ \ \ \ \mbox{} \times
  \left[ [n_{\uparrow}(\varepsilon - \hbar\omega) - n_{\uparrow}(\varepsilon)]
  n_{\downarrow}(\varepsilon - \hbar \Omega) D^{\rm R}(\Omega)
  - [n_{\uparrow}(\varepsilon + \hbar \omega) - n_{\uparrow}(\varepsilon)]
  [1-n_{\downarrow}(\varepsilon - \hbar\Omega)] D^{\rm A}(\Omega) \right]
  \right\} \nonumber \\ 
  =&\,
  i \frac{|J|^2}{\hbar \omega}
  \delta \mu_{\rm s \parallel}(\omega)
  \int d\Omega
  \left\{
  D^{\rm R}(\Omega) \left[\omega [f_{T_{\rm F}}(\Omega - \bar \mu_{\rm m}/\hbar)+1]
    - \int d\varepsilon [n_{\uparrow}(\varepsilon - \hbar \omega) - n_{\uparrow}(\varepsilon)] n_{\downarrow}(\varepsilon - \hbar \Omega) \right]
  \right. \nonumber \\ 
  &\, \left. \ \ \ \ \mbox{}
  + D^{\rm A}(\Omega) \left[(-\omega) [f_{T_{\rm F}}(\Omega - \bar \mu_{\rm m}/\hbar) + 1]
    - \int d\varepsilon [n_{\uparrow}(\varepsilon + \hbar\omega) - n_{\uparrow}(\varepsilon)] n_{\downarrow}(\varepsilon - \hbar \Omega) \right] \right\} \!.
\end{align}
If $T_{\uparrow} = T_{\downarrow} = T$ this may be further simplified as
\begin{align}
  \label{eq:jsresult}
  \delta j^x_{{\rm s}\parallel}(\omega) =&\,
  i \frac{|J|^2}{\hbar \omega} \nu_{\uparrow} \nu_{\downarrow}
  \delta \mu_{\rm s \parallel}(\omega)
  \int d\Omega
  \nonumber \\ &\, \mbox{} \times
  \left\{
  D^{\rm R}(\Omega) \left[(\hbar \Omega- \bar \mu_{\rm s \parallel}- \omega) f_{T_{\rm N}}(\Omega - \omega - \bar \mu_{\rm s \parallel}/\hbar)
    - (\hbar \Omega - \bar \mu_{\rm s \parallel}) f_{T_{\rm N}}(\Omega- \bar \mu_{\rm s \parallel}/\hbar)
    + \hbar \omega f_{T_{\rm F}}(\Omega - \bar \mu_{\rm m}/\hbar)
    \right]
  \right. \nonumber \\ &\, \left. \ \ \ \ \mbox{}
  +
  D^{\rm A}(\Omega) \left[(\hbar \Omega- \bar \mu_{\rm s \parallel} + \hbar \omega) f_{T_{\rm N}}(\Omega + \omega - \bar \mu_{\rm s \parallel}/\hbar)
    - (\hbar \Omega - \bar \mu_{\rm s \parallel}) f_{T_{\rm N}}(\Omega - \bar \mu_{\rm s \parallel}/\hbar)
    - \hbar \omega f_{T_{\rm F}}(\Omega - \bar \mu_{\rm m}/\hbar)
    \right] \right\} \!.
\end{align}
The retarded and advanced magnon Green functions can be obtained from the Krppmers-Kronig relations,
\begin{align}
  D^{\rm R}(\Omega) =&\, D^{\rm A}(\Omega)^* \nonumber \\ =&\,
  \int d\Omega' \frac{\nu_{\rm m}(\Omega')}{\Omega + i \eta - \Omega'},
\end{align}
where $\eta$ is a positive infinitesimal. In the main text the superscript ``R'' for the retarded magnon Green function is omitted.
In the limit $\omega \to 0$, Eq.\ (\ref{eq:jsresult}) simplifies to
\begin{align}
  \delta j^x_{{\rm s}\parallel}(0) =&\,
  2 \pi |J|^2 \nu_{\uparrow} \nu_{\downarrow}
  \delta \mu_{\rm s \parallel}(0)
  \int d\Omega \, \nu_{\rm m}(\Omega)
  \left\{ f_{T_{\rm F}}(\Omega - \bar \mu_{\rm m}/\hbar) - f_{T_{\rm N}} (\Omega - \bar \mu_{\rm s \parallel}/\hbar)
  - (\Omega - \bar \mu_{\rm s \parallel}/\hbar) \left. \frac{d f_{T_{\rm N}}}{d\Omega } \right|_{\Omega - \bar \mu_{\rm s \parallel}/\hbar}\right\} \!,
\end{align}
which is consistent with Eq.\ (\ref{eq:Isdc}).
\end{widetext}

\bibliography{./FFSC_manuscript_arXiv}

\end{document}